\documentclass[aps,twocolumn,floats,pre,nofootinbib,floatfix]{revtex4}
\usepackage{graphics,graphicx,epsfig}
\usepackage{amssymb,color}
\usepackage{amsmath,textgreek,bm}
\usepackage{epsf,epstopdf,wrapfig}
\usepackage{xcolor}
\usepackage{tikz}

\usepackage[colorlinks=true, linkcolor=blue, citecolor=green, urlcolor=magenta]{hyperref}

\def\debug{1}
\newcommand{\Fnotes}[1]{\ifnum\debug=1{\color{purple} [FM: #1]}\fi}

\newcommand{\Cnotes}[1]{\ifnum\debug=1{\color{teal} [FM: #1]}\fi}

\newcommand{\LDC}[1]{\ifnum\debug=1{\color{red} [LDC: #1]}\fi}

\newcommand{\sbold}{\boldsymbol{s}}

\newcommand{\nn}{n}
\newcommand{\mm} {m}

\newcommand{\mean}[1]{ \left \langle {#1} \right \rangle }

\newcommand{\ON}{\mathcal{O} \left (N^{-1} \right) }

\newcommand{\bh}{\boldsymbol{h}}

\date{\today}

\begin{document}
\title{Neural subspaces, minimax entropy, and mean--field theory for networks of neurons}

\author{Luca Di Carlo,$^a$ Francesca Mignacco,$^{a,b}$ Christopher W. Lynn,$^{c}$ and William Bialek$^{a,b}$}
\affiliation{$^a$Joseph Henry Laboratories of Physics and Lewis--Sigler Institute, Princeton University, Princeton NJ 08544 USA}
\affiliation{$^b$Initiative for the Theoretical Sciences, The Graduate Center, City University of New York, 365 Fifth Ave, New York NY 10016 USA}
\affiliation{$^c$Department of Physics, Quantitative Biology Institute, and Wu Tsai Institute, Yale University, New Haven CT 06510 USA}

\date{\today}
\begin{abstract}
Recent advances in experimental techniques enable the simultaneous recording of activity from thousands of neurons in the brain, presenting both an opportunity and a challenge:   to build meaningful, scalable models of large neural populations. 
Correlations in the brain are typically weak but widespread, suggesting
that a mean-field approach might be effective in describing real neural populations, and we explore a hierarchy of maximum entropy models guided by this idea.  We begin with models that match only the mean and variance of the total population activity, and extend to  models that match the experimentally observed mean and variance of activity along multiple projections of the neural state.  Confronted by data from several different brain regions, these models are driven toward a first-order phase transition, characterized by the presence of two nearly degenerate minima in the energy landscape, and this  leads to predictions in qualitative disagreement with other features of the data. To resolve this problem we introduce a novel class of models that constrain the full probability distribution of activity along selected projections. We develop the mean-field theory for this class of models and apply it to recordings from 1000+ neurons in the mouse hippocampus. This ``distributional mean--field'' model provides an accurate and consistent description of the data, offering a scalable and principled approach to modeling complex neural population dynamics.
\end{abstract}

\maketitle
\date{\today}
\section{Introduction} 
\label{sec:intro}
The exploration of brain activity has been revolutionized by the ability to record simultaneously from thousands of neurons \cite{AllenData,gauthier2018dedicated,manley2024simultaneous,Meshulam+Bialek_2024}. The patterns of activity across these large numbers of cells surely are not completely random, but they also are highly variable.  It is natural to think of activity patterns as the microscopic states of the neural network, perhaps mapping to macroscopic states that correlate with or even determine the animal's percepts, plans, and motor actions.  We would like to describe the probability distribution out of which these  patterns are drawn.  In equilibrium statistical mechanics, the distribution over microscopic states---the Boltzmann distribution---contains an enormous amount of information about the system, but extracting this information remains a hard problem.  Being confident that we can write down the distribution over states for a neuronal network similarly would provide a starting point, not an ending point. 

Concretely, we represent each cell with a binary variable $s_{\rm i }$, where $s_{\rm i} = 1$ if the neuron is firing and $s_{\rm i } = -1$ if the neuron is silent in a small window of time; the  state of the network as a whole then is $\sbold \equiv \{s_{\rm i}\}$.  Our goal is to construct the probability distribution $P(\sbold )$, guided by experimental observations. 
Building on two decades of work \cite{Meshulam+Bialek_2024}, we approach this problem using maximum entropy methods \cite{jaynes1957information,jaynes1982rationale}.  In this approach we take seriously the experimental estimates of expectation values for a limited set of observables $\{f_\mu (\sbold ) \}$, insisting that our model reproduce these observations, that is 
\begin{equation}
\langle f_\mu (\sbold ) \rangle_P = \langle f_\mu (\sbold )\rangle_{\rm exp} ,
\label{constraint1}
\end{equation}
where $\mean{\bullet}_P$ and $\mean{\bullet}_{\rm exp}$ are respectively the average over the probability distribution $P(\sbold )$  and the temporal average over the experimental data.  Among all the distributions that satisfy these constraints, we choose the one which has the least structure, so that states drawn from the distribution are as random as possible while still obeying Eq~(\ref{constraint1}).  The search for minimal structure or maximal randomness is (uniquely) mathematized as the distribution with maximum entropy, and this has the form of an energy based model, 
\begin{eqnarray}
    P(\sbold) &=& \frac 1 Z \exp\left[ -E(\sbold) \right]  
    \label{maxent1} \\
    E(\sbold) &=& \sum_\mu g_\mu f_\mu (\sbold ) ;
    \label{maxent2}
\end{eqnarray}
there is a coupling constant $g_\mu$ for each constraint in Eq~(\ref{constraint1}).  Importantly the chosen features of the data determine these coupling constants, so that all subsequent predictions are parameter free.

The functional form of $E(\sbold)$ depends on the observables we have decided to measure, and the choice of the right observables is crucial to the success of the maximum entropy construction. A rather natural choice is the mean activities and the matrix of the pairwise correlations; this choice of observables gives an energy function $E(\sbold)$ that is equivalent to a fully connected spin--glass model with pairwise interactions \cite{mezard+al_87},
\begin{equation}
    E_{\rm pairs} = - \sum_{n} h_{n} s_{n} -{1\over 2} \sum_{\nn \mm} s_\nn J_{\nn \mm} s_{\mm}  ,
    \label{eq:energy}
\end{equation}
where our sign convention is that positive fields $h_n$ favor activity and positive couplings $J_{nm}$ favor simultaneous activity.
This class of models is notably broad; well-known examples include  Boltzmann Machines \cite{ackley+al_85} and Hopfield networks \cite{hopfield1982,hopfield+tank1985,hopfield+tank1986,amit+al_1987}.

This class of pairwise models has been extremely successful in describing many populations of neurons with $N\sim 100$ \cite{Meshulam+Bialek_2024}. However, these models come with certain limitations: constructing them requires access to all $\sim N^2 $ elements of the correlation matrix. While this is not a problem for small neural populations it can become a problem for large ones. Two key factors come into play: the size of the neural population $N$ and the temporal duration of the experimental recordings $T$.  Recent experiments have seen a dramatic increase in $N$ without a corresponding increase in $T$, meaning that the total number of samples $N \times T$ is not sufficient to estimate all $N^2$ pairwise correlations. Even though individual entries $\mean{s_n s_m}_{\rm exp}$ may be estimated accurately, these estimates are not independent. In the extreme case where $T<N$, the correlation matrix is not of full rank. More strongly in writing $P(\sbold )$  we are making the implicit  assumption that the underlying neural activity is stationary,  but circadian rhythms, learning, and representational drift \cite{histed2014cortical, rule2019causes}  restrict the time window over which this assumption holds to just a few hours.

When we are limited by the number of samples, it is still acceptable to ask for experimental estimates of $M\propto N$ expectation values. In this regime, choosing the right observables becomes especially important.  As an example, in  flocks of birds  we can build successful models by choosing to match only the correlations among near neighbors, restricting the effective interactions to be local in space \cite{bialek2012statistical,cavagna+al_2015}. Viewed from a different perspective, models with only ${\cal O } \left (N\right)$ parameters can often generate rich correlation structures that effectively populate the entire $N \times N$ correlation matrix. For instance, an Ising model with nearest neighbor interactions can still produce complex and nontrivial long-range correlations. This suggests that, in principle, building a good model of neural populations with only ${\cal O} \left (N\right)$ parameters should be possible. However, this line of reasoning leaves open an important question: in the absence of symmetry, locality, or conservation laws, how should we choose which observables to constrain?

The literature offers several strategies for reducing the dimensionality of neural data. One  intuition is that the population activity---the average firing rate across the network---captures collective effects arising from neural interactions \cite{duncker2021dynamics}. Related ideas suggest that the high-dimensional dynamics of neural populations is controlled by a small number of latent variables or fields. In this view, the relevant dynamics lie on a low-dimensional manifold, such that a small number of linear, or non-linear, projections of neural activity suffice to characterize the state of the network \cite{Shenoy2013, Cunningham2014,Carsen2019, Edward2021, Recanatesi2021}.

In the maximum entropy framework, the model that matches the mean and variance of the population activity corresponds to a mean-field ferromagnet. More generally, models that match the covariance matrix of a set of $K$ projections of the neural state can be interpreted as generalized mean-field models and are mathematically equivalent to models with latent fields. In this paper we develop the mean--field theory for these models and show their fundamental limitations when applied to real neural populations. We then introduce a novel class of mean--field models that can successfully describe large neural populations.

The remainder of this paper is structured as follows. 
In \S\ref{sec:pop} we revisit the naive mean-field theory applied to population activity. We demonstrate the existence of an upper bound to the fluctuations $\chi$ at fixed mean population activity $\mu$, defining a region in the $\mu$--$\chi$ plane that is inaccessible under the mean-field approximation. 
Remarkably, experimental data across various brain regions, species, and experimental methodologies consistently lie within this forbidden region.
We then solve the inverse Ising problem exactly for neural populations of moderate size, revealing that the maximum entropy solution lies near a first--order phase transition, characterized by switching between low- and high-activity states.  This is inconsistent with experiments, showing that the mean and variance of population activity alone are not sufficient to capture collective effects in these networks.

In \S\ref{sec:projections},we extend the naive mean-field theory to models that match the covariance of fluctuations along multiple projections of neural activity \cite{Cocco2011}. This extension bridges mean-field theory with latent variable models and Hopfield networks.  We solve the corresponding inverse problem within the mean-field approximation. Our findings indicate that models in this class again fail when applied to real data---even at a qualitative level. When they are not trivial, they exhibit issues similar to those of the population activity model. 

Finally, in \S\ref{sec:extended} we introduce a new class of maximum entropy models. These models match the full probability distribution of a projection of neural activity, and are  connected to models for dense  associative memory, or ``modern Hopfield'' networks \cite{krotov+hopfield_2016}. We provide a mean-field solution to the inverse problem and show that these models are consistent and give a good description of real neural populations.

In the background of our discussion are ideas about entropy as a measure of model quality, the way in which this applies to maximum entropy models, and the emergence of the miniMax entropy principle.  These results have a long history, even if some are less well known than they might be.  We give a brief review in Appendix~\ref{app:MaximumEntropy}.

\section{Population activity: mean--field theory and exact solution }
\label{sec:pop}

One of the simplest and most intuitive strategies for dimensionality reduction is to monitor the summed activity, or equivalently the average firing rate, of the neural population. In this section we analyze the maximum entropy model that matches the mean and the variance of this population activity.  This model is mathematically equivalent a fully-connected ferromagnet. We begin by reviewing the textbook solution of the model in the mean--field approximation. Then, we  derive an upper bound on the fluctuations $\chi$ at fixed mean average population activity $\mu$ within this approximation, and show that real neural populations systematically violate  this bound. Finally, we compute the exact solution to the maximum entropy problem and demonstrate that to violate the mean--field bound requires parameters poised close to a first-order phase transition, so that there is a double-well structure in the free energy landscape. 

The population activity is the sum over  all the variables   in the network.  We want to start with a model that matches the (normalized) mean of this activity
\begin{eqnarray}
\mu &=& {1\over N} {\bigg\langle} \left(\sum_n s_n\right) {\bigg\rangle}\\
&=& {1\over N} \sum_n \langle s_n \rangle 
\end{eqnarray}
and its (normalized) variance
\begin{eqnarray}
\chi &=& {1\over N} {\bigg\langle} \left(\sum_n s_n\right)^2 {\bigg\rangle} - 
{1\over N} {\bigg\langle} \left(\sum_n s_n\right) {\bigg\rangle} ^2 \\
&=& {1\over N} \sum_{nm}\langle s_n s_m\rangle^{\rm (c)},
\end{eqnarray}
where $\rm (c)$ denotes the connected part of the correlations.  The variance $\chi$ is equivalent to the magnetic susceptibility in the corresponding models of magnets.
The maximum entropy model that matches these first two moments of the population activity is of the form in Eqs~(\ref{maxent1}, \ref{maxent2}) with the energy function
\begin{equation}
    E_{\rm pop} (\sbold ) = - h \left( \sum_{n }s_n\right)   - \frac \lambda{ 2 N} \left(\sum_n   s_{n } \right)^2 ;
    \label{eq:Epop}
\end{equation}
as usual we insert a factor of $N$ to be sure that both terms in the energy function are of order $N$ (extensive).
The external field $h$ and the coupling constant $\lambda$ are determined by the implicit conditions $\mu_P = \mu_{\rm exp}$ and $\chi_P = \chi_{\rm exp}$.  
Matching these moments can be quite laborious and, in general, involves extensive numerical simulations.  But Eq~(\ref{eq:Epop}) defines a ``mean--field model'' in which every neuron interacts with the average over all other neurons, and at large $N$ this class of models can (usually) be solved analytically.

\subsection{Mean--field solution and  bound in the \texorpdfstring{$\mu$}{mu}--\texorpdfstring{$\chi$}{chi}}

It is a textbook exercise to solve the model defined by Eq~(\ref{eq:Epop}) in the mean--field approximation \cite{parisi1988statistical,sethna2021statistical}. 
The partition function is:   
\begin{equation}
    Z_{\rm pop} = \sum_{\sbold} \exp \left[ h \left(\sum_n  s_n\right)  + \frac \lambda {2N} \left(\sum_n s_n\right)^2\right] , 
    \label{eq:Zpop0} 
\end{equation}
As usual we can obtain expectation values by differentiating the free energy $F = -\ln Z_{\rm pop}$, 
\begin{eqnarray}
\mu &=& -{1\over N}{{\partial F}\over{\partial h}}\\
\chi &=& {1\over N}{{\partial^2 F}\over{\partial h^2}} .
\end{eqnarray}

The partition function be rewritten exactly using the Hubbard--Stratonovich transformation \cite{parisi1988statistical}:
\begin{eqnarray}
Z_{\rm pop} (h,\lambda )&=& \sqrt{\frac N { 2\pi\lambda}} 2^N 
 \int \ d\psi\, e^{-N f(\psi )},
 \label{eq:Zpop_HS}\\
 f(\psi ) &=& \frac{1}{2\lambda}\psi^2 - \ln\cosh(h + \psi) .
\label{eq:fpop}
\end{eqnarray}
At large $N$ the integral in Eq~\eqref{eq:Zpop_HS} should be dominated by the saddle--point $\psi^\star = \lambda \tanh( \psi^\star + h)$ that extremizes the  local free energy $f(\psi )$. This leads to the mean--field free approximation
\begin{equation}
F_{\rm MF} (h,\lambda )=N f(\psi_* ) + N\ln 2 + \frac{1}{2}\ln[2\pi\lambda f''(\psi_*)]  + ~ \cdots ,
\label{lnZ_pop}
\end{equation}
where the ellipsis denotes subleading terms of order $1/N$. 

The field $\psi$ can be interpreted as the effective fluctuating field acting on each neuron and generated by the other neurons.  Approximating the integral with its saddle point is equivalent to replacing $\psi$ with its average value; this is the hallmark of the mean--field approximation. The mean activity $\mu$ and susceptibility $\chi$ can be obtained by differentiating the free energy with respect to $h$. In the same approximation we obtain the self-consistent equations
\begin{eqnarray}
\mu  &=& \tanh (h + \lambda \mu ) +  \ON,
\label{eq:mu_mf}\\
\chi &=& \frac{ 1- \mu^2}{ 1 - \lambda(1- \mu^2)} + \mathcal O \left ( N^{-1} \right )  .
\label{eq:chi_mf}
\end{eqnarray}
These self--consistent equations can be inverted to extract the mean--field solution to the inverse problem. However, before doing so, we must carefully consider the domain of definition of $\mu$ and $\chi$. 

For a fixed population activity $\mu$, the maximum susceptibility $\chi$ is achieved when $\lambda$ is as close as possible to the critical value $\lambda_c = (1 - \mu^2)^{-1}$, while still satisfying Eq~\eqref{eq:mu_mf}. Differentiating Eq~\eqref{eq:mu_mf} implicitly at constant $\mu$ gives:
\begin{equation}
    \left. \frac{d\lambda}{dh} \right|_{\mu} = -\frac{1}{\mu}.
    \label{eq:dldh_mf}
\end{equation}
This implies that increasing the field $h$ at fixed $\mu$ requires reducing the external field $\lambda$. Therefore, the model that yields the largest possible susceptibility at a given $\mu$ corresponds to $h = 0$. Under this condition, Eqs~\eqref{eq:mu_mf} and~\eqref{eq:chi_mf} yield an upper bound on the susceptibility:
\begin{equation}
    \chi_{\mathrm{max}}(\mu) = \frac{ \mu(1 - \mu^2) }{ \mu - \mathrm{atanh}(\mu)(1 - \mu^2) }.
    \label{eq:nmf_chimax}
\end{equation}
This relation defines the boundary of the region in the $\mu$--$\chi$ plane that is accessible under the mean-field approximation. 
When the empirical moments $\mu$ and $\chi$ obey this bound, we can use Eqs~\eqref{eq:mu_mf} and~\eqref{eq:chi_mf} to set the predicted moments equal to their experimental values, and the equations can be solved to give
\begin{align}
    \lambda_{\rm MF} &= \frac{1}{1 - \mu_{\rm exp}^2} - \frac{1}{\chi_{\rm exp}} + \mathcal{O}(1/N),
    \label{eq:lambda_mf} \\
    h_{\rm MF} &= \mathrm{atanh}(\mu_{\rm exp}) - \lambda_{\rm MF} \mu_{\rm exp} + \mathcal{O}(1/N).
    \label{eq:h_mf}
\end{align}

\begin{figure}
    \includegraphics[width=0.95\linewidth]{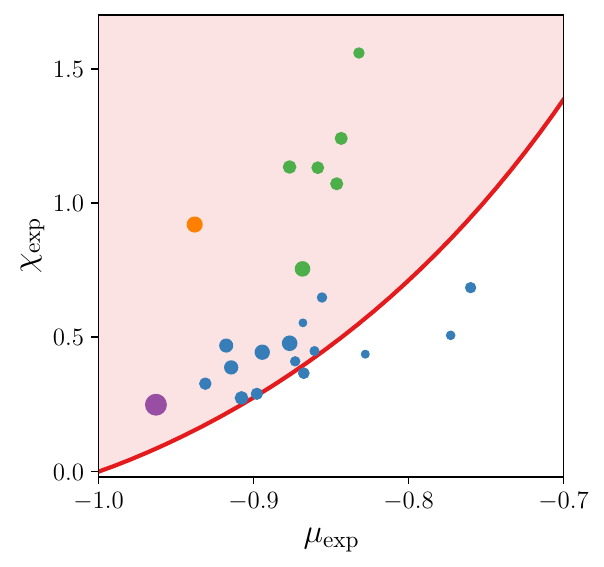}
    \caption{
        Susceptibility $\chi$ vs mean activity $\mu$. Equation~(\ref{eq:nmf_chimax}) defines an upper bound $\chi_{\rm max}(\mu)$ (red line); values $\chi > \chi_{\rm max}$ (shaded) cannot be found in the mean--field approximation.
Data are from multiple experiments: $N = 160$ neurons in the retina (orange), recorded with an electrode array~\cite{tkavcik2014searching}; $N = 1416$ neurons in the mouse hippocampus (purple), recorded via calcium imaging~\cite{gauthier2018dedicated,meshulam2019RG}; $N = 60 - 190$ neurons from single brain regions (blue) and $N = 900 - 1400$ across multiple regions (green), both recorded using Neuropixels 2.0 \cite{AllenData}.  Symbol sizes reflect the value of $N$ in each case.
 \label{fig:chi_mu_data}}
\end{figure}

Perhaps surprisingly we find that large neuronal populations consistently violate the bound  in Eq~(\ref{eq:nmf_chimax}), and this is true across  a wide range of brain regions, species, and experimental modalities. In Figure~\ref{fig:chi_mu_data} we plot the experimental values of susceptibility $\chi_{\rm exp}$ versus the mean activity $\mu_{\rm exp}$ for several datasets, comparing them against the  bound. Experiments include $N=160$ output neurons from the vertebrate retina, responsible for transmitting visual information from the eye to the brain \cite{tkavcik2014searching}; populations of $N \sim 60 -190$ neurons across various regions of the mouse brain, such as the visual and motor cortices and the hippocampus \cite{AllenData}; $N\sim 900 - 1500$ neurons across multiple mouse brain areas \cite{AllenData}; and larger-scale recordings of $N \sim 1400$ neurons in the CA1 region of the mouse hippocampus \cite{gauthier2018dedicated,meshulam2019RG}.

The solution to the inverse problem defined by Eqs~\eqref{eq:lambda_mf} and~\eqref{eq:h_mf} appears agnostic as to whether $\chi_{\rm exp}$ and $\mu_{\rm exp}$ are consistent with the bound. In fact, the mean--field estimates $\lambda_{\rm MF}$ and $h_{\rm MF}$ can be computed for {\em any} empirical values of $\mu_{\rm exp}$ and $\chi_{\rm exp}$, and one might be tempted to apply these formulas directly to  data in the hope of recovering meaningful parameters. However, doing so leads to qualitatively incorrect predictions: for instance, in cases where the bound is violated, the inferred external field $h_{\rm MF}$ often has the opposite sign of the empirical mean activity $\mu$. Furthermore, inserting the mean--field solution of the inverse problem back into the mean-field equations, for pairs $\mu$--$\chi$ outside of the bound, leads to a contradiction: $\mu_{\rm exp} \neq \tanh(h_{\rm MF} + \lambda_{\rm MF} \mu_{\rm exp})$. This inconsistency reveals that the inferred parameters $\lambda_{\rm MF}$ and $h_{\rm MF}$ are incorrect. The root of the issue lies in having inverted Eqs~\eqref{eq:mu_mf} and~\eqref{eq:chi_mf} outside their domain of validity.

\subsection{Exact solution of the mean--field model} 

The observation that real neuronal populations sit in the region of the $\mu$--$\chi$ plane that is inaccessible to the mean--field approximation raises an important question: where is the solution to the maximum entropy problem? It seems reasonable to assume that a maximum entropy distribution must still exist---among all distributions consistent with the observed moments, there is one that has the maximum entropy.
But we do know of cases in which  the maximum entropy distribution  sits on an edge of the space of probabilities \cite{DowsonD.1973,Tagliani2003,Inverardi2021}, so that the distribution with the highest entropy is not a stationary point $\delta S / \delta P = 0$. A more careful analysis  is needed.  

\begin{figure*}[t]
    \includegraphics[width=\linewidth]{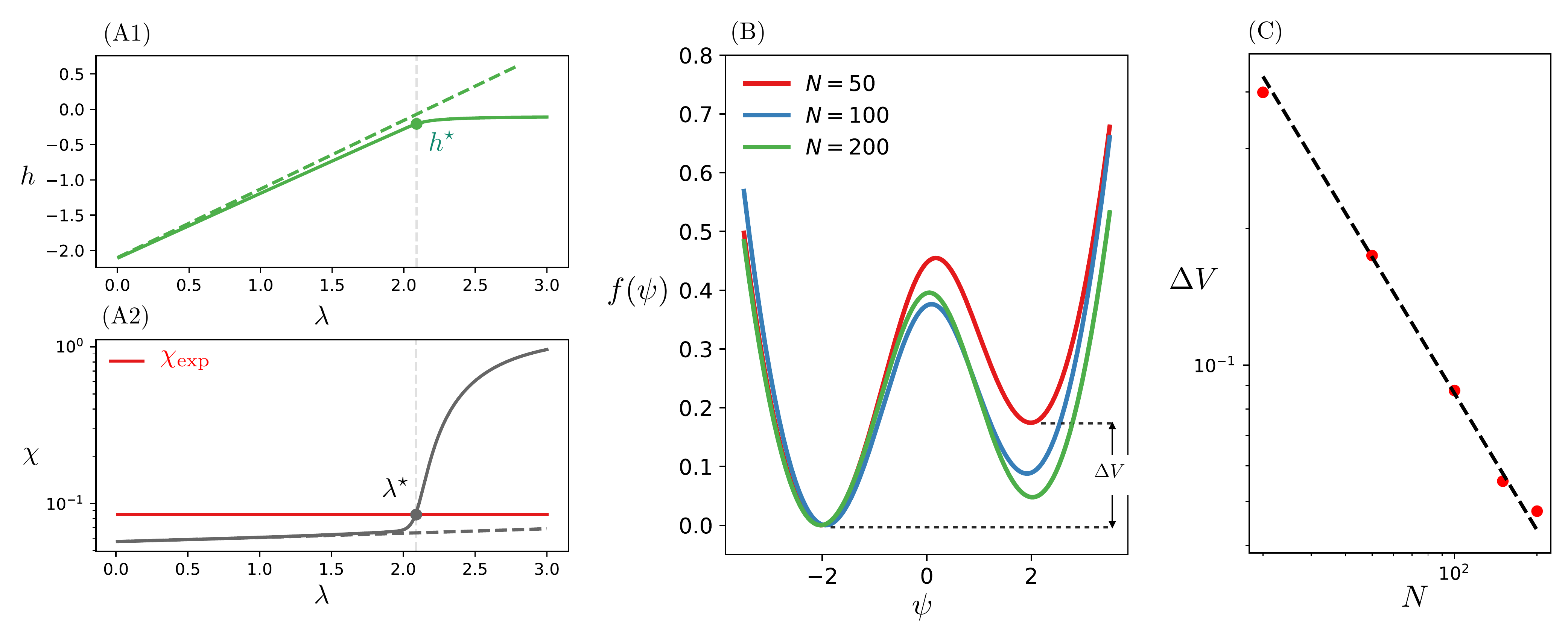}
    \caption{Exact solutions of the mean--field model. (A) Trajectories of $h(\lambda)$  and the corresponding susceptibility  $\chi(\lambda)$ at constant magnetization $\mu_{\rm exp}$ obtained via exact integration (solid lines) and with the mean--field approximation (dashed lines) for a population of $N=50$ neurons from the hippocampus. For large enough $\lambda$ the mean--field approximation prediction  deviates from the exact solution.  The intersection between the experimental susceptibility (red line) and $\chi(\lambda)$ (gray curve) determines the exact solution $\lambda ^\star$ and $h^\star$ of the maximum entropy problem. The experimental variance is such that the solution is outside of the regime of validity of the mean--field approximation. (B) Local free energy per neuron $f(\psi )$ with parameters $h$ and $\lambda$ inferred from experiments  on a network of neurons in the mouse hippocampus \cite{gauthier2018dedicated,meshulam2019RG}.  As we consider larger populations (increasing $N$), the two local minima become more nearly degenerate, signaling the proximity of a first order phase transition. (C) Local free energy difference between the two nearly degenerate minima for neural populations of increasing size $N$. The energy difference $\Delta V$ goes to zero as the system size increases and it is well described by $\Delta V \sim N^{-1}$. \label{fig:exact}}
\end{figure*}

Having identified the failure of the mean--field approximation, we now turn to the exact solution of the model. For populations of moderate size $N$ we can compute the partition function $Z_{\rm pop}(h, \lambda)$ exactly by numerically integrating Eq~(\ref{eq:Zpop_HS}).   We also have exact equations for $\mu$ and $\chi$,
\begin{eqnarray}
    \mu  &=& \frac{1}{ Z_{\rm pop}} \sqrt{\frac N { 2\pi\lambda}} 2^N  \int  \tanh(\psi + h) e^{-N f(\psi)} d\psi 
    \label{eq:exactMu}\\   
        \chi  &=& 1 - N \mu ^2 \nonumber\\
        &&\,\,\,\,\,+   \frac{(N-1) \sqrt{N} 2^N }{Z_{\rm pop} \sqrt{2\pi\lambda}}  \int  \tanh^2 (\psi + h) e^{-N f(\psi)} d\psi  \nonumber\\
        &&
    \label{eq:exactChi}
\end{eqnarray}
These equations give an explicit solution to the direct Ising problem, but it is not evident how to invert them to obtain $h(\mu, \chi) $ and $\lambda(\mu, \chi) $. We can solve this problem by taking inspiration from the characteristics method used to solve ordinary differential equations \cite{evans2022partial}.  

The experimental mean activity $\mu_{\rm exp}$ defines a one dimensional manifold in the space of parameters, $h = h(\lambda)$, of all the models that satisfy $\mu = \mu_{\rm exp}$. Changing the value of the parameters by $dh$ and $d\lambda$ changes the predicted value of the population activity by
\begin{equation}
    d \mu = \frac{ \partial \mu}{ \partial \lambda }   d \lambda  + \frac{\partial \mu}{ \partial h} d h
\end{equation}
Therefore, we can surf the constant activity manifold by solving the differential equation
\begin{equation}
    \frac {dh(\lambda)}{ d \lambda} = - \left(\frac{\partial \mu}{ \partial h}\right) ^{-1}  \frac{ \partial \mu}{ \partial \lambda}.
\end{equation}

In Figure \ref{fig:exact} we show how this procedure works for a population of neurons in the hippocampus, starting with $N=50$.  The trajectory $h(\lambda)$ starts from $h(0) = \operatorname{atanh}( \mu)$ and, in agreement with the mean--field approximation, Eq~\eqref{eq:dldh_mf}, it increases linearly towards $h=0$ as $\lambda$ increases (Fig~\ref{fig:exact}A1). But just before reaching $h=0$ the mean--field approximation starts to fail and the trajectory stalls, so that $dh/ d\lambda$ is almost zero, and this corresponds to a rapid rise of the susceptibility. The experimental susceptibility intersects $\chi(\lambda)$ on this steep rise and $\lambda^\star$ is determined very precisely (Fig~\ref{fig:exact}A2).

After we have found the parameters $h^\star$ and $\lambda^\star$ that match the experimental moments, we can plot the  local free energy  $f(\psi)$ from Eq~\eqref{eq:fpop}. The local free energy has two nearly degenerate minima, as seen in Fig~\ref{fig:exact}B.  This provides a hint as to why the mean--field approximation is breaking down:  with two local minima of $f(\psi)$ the integral that defines the partition function in Eq~(\ref{eq:Zpop_HS}) has two saddle points rather than one, and if the difference in value of the local free energy between these two points is $\sim 1/N$ then both will contribute even in the $N\rightarrow\infty$ limit \cite{bender2013advanced}.  To see if this happens we need access to a population of neurons where we can let $N$ vary systematically.

Optical imaging experiments on the CA1 region of the mouse hippocampus record from a population of $1000+$ neurons that are in a single plane \cite{gauthier2018dedicated,meshulam2019RG}, and so it makes sense to change the size of the population that we analyze by changing the radius of a circle inside the field of view \cite{Meshulam2021,Lynn2023,Meshulam+Bialek_2024}.  Figure \ref{fig:exact}B shows that the two minima persist as we increase from $N=50$ to $N=200$, and indeed the gap between the minima varies in proportion to $1/N$ (Fig~\ref{fig:exact}C).  This explains the breakdown of the mean--field approximation.

\subsection{Conclusions}

The results of this section highlight  the important distinction between a {\em mean--field model} and the {\em mean--field approximation}.   A mean--field model, like the one defined in Eq~\eqref{eq:Epop}, is characterized by interactions that couple each neuron or spin to an average of all the other variables.  We can have mean--field ferromagnets, as in the present case, or mean--field spin glasses where the averaging involves random weights \cite{mezard+al_87}.   In contrast, the mean--field approximation is a technique used to compute the partition function using the saddle-point method \cite{parisi1988statistical}. While it is generally reasonable for mean--field models to be solvable in the mean--field approximation at large $N$, this is not always the case.  In the context of maximum entropy models,  the data that decide whether we are in a regime whether the mean--field approximation is applicable.

The distinction between mean--field models and the mean--field approximation proves to be essential in building the least structured model that matches the mean and variance of activity in an entire neural population.   We found this surprising.  Unfortunately now that we can complete the maximum entropy construction we see that the resulting model is a very bad description of the data.  Because the local free energy $f(\psi)$ has two near--degenerate minima, the model predicts that the distribution of summed activity will be bimodal, with the whole network switching synchronously between highly active and nearly silent states.  This is not what we see in the experiments.   The conclusion is that collective activity in these networks cannot be captured just by matching the mean and variance of the summed population activity---we need more structure. 

\section{Models Constraining Multiple Projections}
\label{sec:projections}

The failure of the simplest population activity model suggests that we need more structured models if we want to describe real neural populations. After all, the summed activity is only one particular projection of the full neural state. We could instead consider different projections, or even multiple projections simultaneously.  Here we extend the mean--field theory to models that constrain the variance of multiple projections of the neural state. These models capture richer correlation structures and relate to classical models of associative memory \cite{amit+al_1987}. We derive the corresponding mean--field equations and show that, for randomly chosen projections, the mean--field approximation is consistent when applied to experimental data but provides little information about the system. To address this limitation, we identify optimal projections by solving the miniMax entropy principle \cite{zhu+al_1997,MEMysteries,grendar2001minimax,Lynn2023,Lynn+al_2025b}, as discussed in Appendix \ref{app:MaximumEntropy}. Finally, we demonstrate that even optimal projection models suffer from the same fundamental issues identified in the simple population activity model.

\subsection{Formulating and solving the models}

The model discussed above is limited in two ways.  First, by considering only the summed population activity all individuality of the neurons is lost; we even miss the fact that different neurons in the network have different mean activities.  Second, the choice of summed activity is a very specific example of dimensionality reduction, and is very restrictive.  

To go beyond these limitations we want  a model that matches the experimentally observed mean activity of each neuron, 
\begin{equation}
\mu_n = \langle s_n \rangle .
\end{equation}
In addition, we consider projections of the activity,
\begin{equation}
     \varphi_\alpha = \frac{1}{\sqrt{N}}\sum_{n=1}^N W_{\alpha n} s_n, \qquad \alpha = 1, \dots, K , 
\end{equation}
and ask that the model match the experimentally observed covariance along these projections,
\begin{equation}
\chi_{\alpha\beta} = \langle \varphi_\alpha \varphi_\beta\rangle^{\rm (c)} ,
\end{equation}
where again $\rm (c)$ denotes the connected part.
While the performance of the model will inevitably depend on the choice of the projections $W_{\alpha n}$, the overall theoretical framework remains independent of this choice. We therefore leave the choice of projections unspecified for now.  The model that matches these quantities has the Boltzmann form in Eq~(\ref{maxent1}), with the energy function
\begin{equation}
 E_{\rm proj}(\mathbf{s}) = -\sum_{n=1}^N h_n s_n -\frac{1}{2N}\sum_{\alpha, \beta, n, m} s_n W^T_{n \alpha} \Lambda_{\alpha \beta} W_{\beta m} s_m . \label{eq:Eproj} 
 \end{equation} 
Models of this type are reminiscent of Hopfield networks for associative memory \cite{hopfield1982}, where the projections  are analogous to the stored patterns, although this mapping is not exact \cite{amit+al_1987,Cocco2011}. 

The  mean--field approximation and its use to solve the inverse problem follow  the same steps as for the simpler population activity model described in \S\ref{sec:pop}. We first derive an integral representation for the partition function: 
\begin{equation} 
    Z_{\rm proj} = \sqrt{N\over{(2\pi)^K|\Lambda |}}\int  {d^K \psi}  \,\exp\left[ -N f_{\rm proj}(\boldsymbol{\psi}) \right], 
    \label{eq:Zproj}
\end{equation} 
where $|\Lambda|$ is the determinant of the matrix $\Lambda$ and the local free energy \begin{widetext}
\begin{equation} 
    f_{\rm proj}(\boldsymbol{\psi}) = \frac{1}{2} \boldsymbol{\psi}^T \Lambda^{-1} \boldsymbol{\psi}     - \frac{1}{N} \sum_{n=1}^N \ln \cosh\left( h_n + \sum_{\alpha=1}^K W_{\alpha n} \psi_\alpha \right) .
    \label{eq:fproj} 
\end{equation}
\end{widetext}

In the mean-field approximation, the partition function is evaluated by saddle-point integration, leading to the free energy
\begin{equation}
    F(h_n, \Lambda_{\alpha \beta}) = N f_{\rm proj} - N \ln 2 + (\boldsymbol{\psi}^\star) + \frac{1}{2} \ln \left| \mathbb{I} - \Lambda \Delta \right| \dots,
    \label{eq:projection_freeEnergy}
 \end{equation} 
 where the components of the saddle point vector $\boldsymbol{\psi}^\star$ obey  \begin{equation} 
    {\psi}^\star_\alpha = \frac{1}{N}\sum_{\beta = 1}^K \Lambda_{\alpha\beta} \sum_{n=1}^NW_{\beta n}\tanh\left(h_n + \sum_{\gamma = 1}^K W_{\gamma n} {\psi}^\star_\gamma \right) 
    \label{eq:projection_SP}
 \end{equation} 
 and the matrix $\Delta$ has elements 
 \begin{eqnarray} 
    \Delta_{\alpha \beta} &=& \frac{1}{N} \sum_{n=1}^N W_{\alpha n} 
    \left[ 1 - \left(\mu_n^{(0)}\right)^2\right] W_{\beta n} , 
    \label{Delta1}\\
    \mu_n^{(0)} &=& 
    \tanh\left( h_n + \sum_{\gamma=1}^K W_{\gamma n} \psi^\star_\gamma \right) .
    \label{mu0}
\end{eqnarray} 
At the leading order $\Delta_{\alpha\beta}$ coincides with the covariance matrix of the projections that we would see if the neurons were independent, but with their observed mean activities.  This has a non-trivial structure because each neuron contributes simultaneously to multiple projections.

The mean activities $\mu_n = \langle s_n \rangle$ and the covariance 
\begin{equation}
C_{nm} = \langle s_n s_m \rangle - \langle s_n \rangle\langle s_m \rangle
\label{C_def}
\end{equation}
are given as usual by derivatives of the free energy: 
\begin{eqnarray} 
\mu_n &=& -\frac{\partial F( \boldsymbol{h} , \Lambda)}{\partial h_n}, \\
C_{nm} &=& -\frac{\partial^2F(\boldsymbol{h},  \Lambda)}{\partial h_n \partial h_m}, 
\label{eq:diconnectedChiPred} 
\end{eqnarray} 
The connected correlations between the projections are then 
$\chi = WCW^T$.
In the mean-field approximation these quantities become,
\begin{eqnarray}
     \boldsymbol{\mu} &=& \tanh\left( \boldsymbol{h} + W^T \boldsymbol{\psi}^\star \right) + \frac{ 1}{ N} \boldsymbol {r} +  \mathcal{O}(1/N^2), \label{eq:projections_mu} \\
     \chi &=& \left( \Delta^{-1} - \Lambda \right)^{-1} + \mathcal{O}(1/N). \label{eq:projections_chi}
 \end{eqnarray}
Corrections of order $1/N^2$,  as well as the small term $\boldsymbol r$, whose derivation can be found in the Appendix \ref{app:corrections},  are negligible for our purposes and are omitted here; thus $\mu_n = \mu_n^{(0)}$ from Eq~(\ref{mu0}).

The solution of the inverse problem is obtained by inverting these equations, with $\boldsymbol{\mu} =\boldsymbol{\mu}_{\rm exp}$ and $\chi = \chi_{\rm exp}$.  The  result is
\begin{eqnarray}
     \Lambda_{\rm MF}  &=& \left( \Delta_{\rm exp} ^{-1} \chi_{\rm exp} - \mathbb{I}  \right) \chi_{\rm exp} ^{-1} +\cdots , \label{eq:projections_Lambda} \\
      \boldsymbol{h}_{\rm MF}  &=& \mathrm{atanh}(\boldsymbol{\mu}_{\rm exp}) - \frac{1}{N} W^T \Lambda _{\rm MF} W \boldsymbol{\mu}_{\rm exp} +\cdots, \label{eq:projections_h} 
\end{eqnarray}
where the dots indicate small $\mathcal{O}(1/N)$ corrections and $\Delta_{\rm exp}$ is what we get by substituting $\boldsymbol{\mu}^{(0)} = \boldsymbol{\mu} = \boldsymbol{\mu}_{\rm exp}$ into Eq~(\ref{Delta1}). 

The entropy of the model is given by $S = \mean{E} - F$, where the mean energy 
\begin{eqnarray}
    \mean{E} = - \bh^T \boldsymbol{\mu}_{\rm exp} - \frac  12 \operatorname{Tr}\left[\Lambda \chi + \Lambda  W \boldsymbol{\mu}_{\rm exp}  \boldsymbol{\mu}_{\rm exp}^T W^T \right] .
\end{eqnarray}
Combining this equation with Eq~\eqref{eq:projection_freeEnergy}, the entropy can be expressed as $S = S_0 - \Delta S$,  where $S_0$ is the entropy of a model that matches the observed mean activities $\boldsymbol{\mu}_{\rm exp}$ of the neurons but correlations are absent,  and the entropy reduction 
\begin{equation} 
    \Delta S(W) = \frac{1}{2} \mathrm{Tr} \left[ Q - \ln(Q) - \mathbb{I} \right]  
    \label{eq:proj_DS} 
        \end{equation}
is the information gained by matching the correlations of the $K$ projections. The matrix $Q$ is given by
\begin{equation}
    Q = \Delta_{\rm exp}^{-1} \chi_{\rm exp} 
    \label{Qdef}
    \end{equation}
in terms of the experimental observables, as derived in Appendix~\ref{app:entropy}.

\subsection{Attractive and repulsive patterns}

The solution to the maximum entropy problem described by Eqs~\eqref{eq:projections_Lambda} and \eqref{eq:projections_h} has a natural interpretation in terms of patterns in a Hopfield-like network. In this context, the columns of the matrix $W$ can be interpreted as patterns in the network, even though strictly speaking these patterns should be binary.

To simplify the discussion and avoid unnecessary complications, let us first consider a model that matches only one projection. In this case the matrix $W$ reduces to a single vector $W_n$, and $\Lambda$ and $\Delta_{\rm \exp}$   both  become scalars. When comparing the variance $\chi_{\rm exp}$ of the projection $\varphi = \sum_n W_n s_n$ with what would be observed in a model of independent neurons $\Delta_{\rm exp}$  we observe the following: if the variance  $\chi_{\rm exp}$ is larger than the independent model prediction, then the coupling constant  $\Lambda$ is positive;  if $\chi_{\rm exp}$ is smaller, the coupling $\Lambda$ is negative.  
This simple result raises two important questions: What is the interpretation of a negative coupling?  And what happens to the stability of the potential described by Eq~\eqref{eq:fproj} when $\Lambda$ is negative?

\begin{figure*}[t]
    \includegraphics[width = \textwidth]{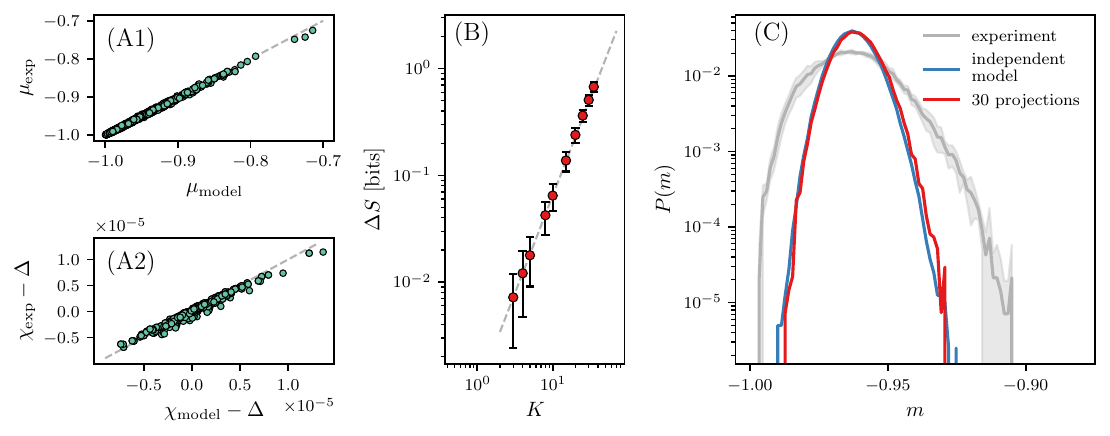}
    \caption{Performance of the maximum entropy model matching mean activities and the covariance of random projections.
    (A) Comparison between model predictions and experimental data for the mean activities $\mu_n$ (A1) and for elements of the covariance matrix relative to the the independent model, $\Delta \chi = \chi - \Delta$ (A2), for a neural population of size $N = 1416$ and $K=30$ random projections. 
    (B) Entropy reduction $\Delta S = S_0 - S$ vs the number of  projections $K$. Points represent the average over multiple random realizations of $W$, with error bars indicating the standard deviation. The dashed line is  $\Delta S = A K^\alpha$ with $\alpha = 1.83(1)$ and $A = 9.6(1)\times 10^{-4}$. Even for $K=30$, the entropy reduction remains small compared to the entropy of the independent model $S_0 \approx 182 \,{\rm bits}$.
    (C) Distribution of the population activity: comparison between experimental data, the maximum entropy model with random projections, and the independent model. Random projections fail to capture the structure of the population activity, and the model predictions remain close to those of the independent model.    \label{fig:randomprojection}
    }
\end{figure*}

From Equation~\eqref{eq:Eproj}, the sign of $\Lambda$ determines the relationship between the projection of the neural state onto the pattern $W_n$ and the variation in energy. For positive $\Lambda$, a larger projection   corresponds to a lower energy, implying that the system favors configurations that align with the vector $W_n$; we could summarize this by saying that $W$ is an attractive pattern.  In contrast, for negative $\Lambda$, a larger projection leads to higher energy, so that the system resists alignment with the pattern $W_n$; we can summarize this by saying that $W$ is a repulsive pattern.
Note that models with negative coupling $\Lambda$ are mathematically allowed, since energy of the model Eq~(\ref{eq:Eproj}) remains bounded regardless of the sign of the interaction. In this case, we can make sense of Eq~(\ref{eq:Zproj})---which would naively yield a divergent result for negative $\Lambda$---by analytic continuation \cite{bender2013advanced}.

\subsection{Random projection models } 
\label{sec:randproj}

How good is the solution we have obtained? This really  is two questions:  is the mean-field approximation consistent, and does the resulting model provide a good description of the data? The model parameters were inferred by matching the experimental quantities $\boldsymbol{\mu}_{\rm exp}$ and $\chi_{\rm exp}$ to their mean--field predictions. However, the mean--field solution is not the exact solution of the maximum entropy problem. To assess the quality of the approximation, we analyze experimental data on $N=1000+$ neurons in the mouse hippocampus \cite{gauthier2018dedicated,meshulam2019RG}\footnote{We neglected all neurons with an average activity below $2 \times 10^{-3}$; this corresponds to neurons firing, on average, fewer than 3 times per minute. } using the mean--field approximation and then simulate the resulting model with Monte Carlo \cite{newman1999monte}.  To the extent that the  mean--field approximation is valid, then when we compute   $\boldsymbol{\mu}_{\rm model}$ and $\chi_{\rm model}$ by averaging over the Monte Carlo samples  we will recover the experimental results $\boldsymbol{\mu}_{\rm exp}$ and $\chi_{\rm exp}$.  If this doesn't work it signals the failure of the mean--field approximation.

We consider a set of random projections with weights chosen as Gaussian random numbers, $W_{\alpha n} \sim \mathcal{N}(0, 1)$. In Figure~\ref{fig:randomprojection}A  we show that fitting the covariance of $K=30$ random projections leads to a model that reproduces reasonably well both the individual activities $\boldsymbol{\mu}_{\rm exp}$ and the elements of the matrix $ \chi_{\rm exp} - \Delta_{\rm exp}$ that measures the deviation of the correlations from the predictions of an independent model.  This confirms that the mean--field approximation is consistent. But whether this results in a good model of the data is a separate question.

As discussed in Appendix \ref{app:MaximumEntropy}, for maximum entropy models the entropy of the model itself is a measure of its quality.  In maximum entropy models that match the covariance of fluctuations along some set of projections, the mean--field approximation relates the entropy reduction $\Delta S$ to measured quantities through Eqs~(\ref{eq:proj_DS}, \ref{Qdef}), and we can think of this as the information that we gain about the states of the network by knowing the covariance of projections.  In Figure~\ref{fig:randomprojection}B we show that the entropy reduction grows with the number of projections approximately as $\Delta S \sim A K^{1.83}$, with a small prefactor $A \sim 9.6(1)\times 10^{-4}$. Even for $K=30$, the entropy reduction remains negligible compared to the entropy of the independent model, $S_0 \approx 128 \, {\rm bits}$. Furthermore it is possible to show analytically that the entropy reduction by matching the variance of a single a random projection scales with the population size $\Delta S \sim 1/N$; see Appendix \ref{app:rand}. We conclude that a model matching the covariance of activity along random projections does not tell us much about the population.

Another way to assess the quality of the model is to evaluate its ability to reproduce observables that were not explicitly constrained by the maximum entropy construction. A simple example is the distribution of the population activity. In Figure~\ref{fig:randomprojection}C, we show that the model matching random projections fails to reproduce the experimental distribution of the population activity, and its predictions are almost indistinguishable from those of the independent model.

We conclude that models based on random projections are ineffective. The poor performance of models based on random projections is not entirely surprising. In high-dimensional spaces, most directions are generically uninformative: random projections are unlikely to align with meaningful collective modes or structured patterns in the data. As a result, models based on such projections fail to capture relevant features of the neural activity. This observation highlights the importance of selecting projections more carefully. 

\subsection{Optimal projections }
\label{sec:optimalProjections}

The energy function of the projection model defined by Equation (\ref{eq:Eproj}) can be interpreted as a fully connected Ising model with an interaction matrix $J_{nm}$ that is constrained  to be of rank $K$.  Identifying the optimal projections then amounts to solving a (challenging!) maximum likelihood problem for an Ising model with a rank-$K$ interaction matrix, from which the projections can then be extracted via a singular value decomposition.
An alternative approach, as motivated in Appendix \ref{app:MaximumEntropy}, is to solve the so-called miniMax entropy problem: construct the maximum entropy model that matches the statistics of $K$ projections and then find the projections $W^\star$ that yield the model with the lowest possible entropy.  These two formulations are  equivalent~\cite{MEMysteries,grendar2001minimax,carcamo+al_2025}, and here we adopt the latter.

\subsubsection{Gauge invariance and principal components}
\label{sec:gauge1}

If $A$ is an invertible matrix, then the energy function in Equation (\ref{eq:Eproj}) is invariant under the transformation 
\begin{eqnarray}
W_{\alpha n} &\rightarrow& \sum_{\beta =1}^K \left( A^{-1} \right)_{\alpha\beta} W_{\beta n}\\
\Lambda_{\alpha\beta} &\rightarrow& \sum_{\gamma =1}^K\sum_{\delta =1}^K  A_{\gamma\alpha} \Lambda_{\gamma\delta} A_{\delta\beta} ,
\end{eqnarray}
or more compactly
\begin{equation} 
    W \to A^{-1} W \quad,\quad \Lambda \to A^T \Lambda A .
\end{equation}
This symmetry implies that a change of basis in the projection matrix $W$ can be absorbed by a congruent transformation of the coupling matrix $\Lambda$---a freedom analogous to gauge invariance in other physics problems. Fixing this gauge appropriately simplifies computation.
 
In particular,  because the matrix $\Delta$ in Eq (\ref{Delta1}) is positive definite we can choose a gauge in which 
$\Delta = \mathbb I $.\footnote{This follows from Sylvester's law of inertia \cite{lang1987linear}.}  To be a bit more explicit, from Eq~(\ref{Delta1}) we can see that $\Delta = \mathbb I $ if the vectors
\begin{equation}
u_{\alpha n} = \sqrt{1-\left( \mu_n^{(0)}\right)^2} W_{\alpha n}
\label{u-W}
\end{equation}
form an orthonormal set
\begin{equation}
{1\over N} \sum_{n=1}^N u_{\alpha n}u_{\beta n} = \delta_{\alpha\beta}.
\end{equation}
We note that the susceptibility
\begin{equation}
\chi_{\alpha\beta} = \sum_{n,m =1}^N W_{\alpha n} C_{nm} W_{\beta m},
\end{equation}
with $C_{nm}$ from Eq~(\ref{C_def}), then becomes
\begin{equation}
\chi_{\alpha\beta} = \sum_{n,m =1}^N u_{\alpha n} {\tilde C}_{nm} u_{\beta m},
\end{equation}
where $\tilde{C}$ is the matrix of correlation coefficients, 
    \begin{equation} 
        \tilde{C}_{nm} = \frac{ \mean{s_n s_m} - \mean{s_n} \mean{s_m} }{ \sqrt{1 - \mean{s_n}^2} \sqrt{1 - \mean{s_m}^2} } \quad .
     \end{equation}
If we chose $u_{\alpha n}$ as eigenvectors of the correlation matrix,
\begin{equation}
\sum_{m=1}^N {\tilde C}_{nm} u_{\alpha m} = \rho_{\alpha} u_{\alpha n} ,
\end{equation}
then $\chi$ takes an especially simple form
\begin{equation}
\chi_{\alpha\beta} = \delta_{\alpha\beta} \rho_\alpha .
\end{equation}
When we substitute into Eqs~(\ref{eq:proj_DS}, \ref{Qdef}) to compute the entropy we find
\begin{equation}
     \Delta S = \frac{1}{2} \sum_{\alpha = 1}^K \left[ \rho_\alpha - \ln(\rho_\alpha) - 1 \right] .
    \label{eq:DS_multiPC} 
\end{equation}
Here and in further results below we give the entropy in nats, as we would in conventional statistical mechanics problems; to obtain the result in bits, divide by $\ln 2$.
Finally, we note that if $u_{\alpha n}$ is a eigenvector of the correlation matrix, then the corresponding $W_{\alpha n}$ from Eq~(\ref{u-W}) is an eigenvector of the covariance matrix.  Thus, it is natural to choose the basis in which the projections $\varphi_\alpha$ are the principal components.

If we focus on a single principal component we find, in agreement with the previous literature \cite{Cocco2011, cocco2018statistical}, \begin{equation}
     \Delta S = \frac{1}{2} \left[ \rho - \ln(\rho) - 1 \right]  . \label{eq:DS_singlePC}
\end{equation}
Two limiting cases can be identified corresponding to large entropy reduction.  First,  when $\rho \gg 1$, the linear term is large. These correspond to high-variance modes, intuitively expected to be informative.  Second,  when $\rho \ll 1$, the $-\ln(\rho)$ term dominates, indicating pseudo-constraints or nearly conserved quantities. These correspond to repulsive patterns that the model avoids.

Solving the miniMax entropy problem (Appendix \ref{app:MaximumEntropy}) reduces to selecting the $K$ components that maximize the sum in Eq~(\ref{eq:DS_multiPC}). This can be done by ranking the components by their individual $\Delta S$~\eqref{eq:DS_singlePC} and picking the first $K$ components.  Interestingly, the eigenvalue spectrum of real neuronal populations contains both very large and very small values (Fig~\ref{fig:projections_DS}A), implying that repulsive patterns might be included in the solution of the miniMax problem when considering many projections.

\subsubsection{Consistency and breakdown of the mean--field approximation}

\begin{figure}
    \includegraphics[width=\linewidth]{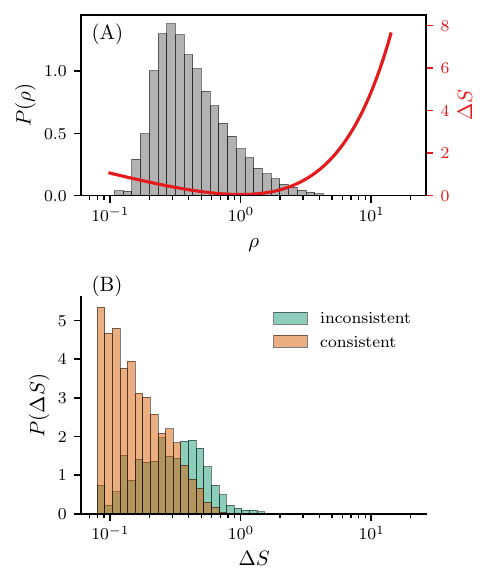}
    \caption{Eigenvalues, entropy, and consistency in models that match the variance of single principal components. (A) Spectrum of the correlation coefficient matrix (grey) and corresponding entropy reduction (red), computed according to Eq~\eqref{eq:DS_singlePC}. 
    (B) Distribution of entropy reductions $\Delta S$ for individual principal components. Orange indicates components for which a consistent mean-field solution exists; cyan indicates components for which the mean-field solution is inconsistent. Principal components compatible with a mean-field description are associated with very small entropy reductions, except for a single outlier; principal components corresponding to large entropy reductions are inconsistent with a mean--field description. Results shown are for a population of $N = 1416$ neurons recorded from the hippocampus \cite{gauthier2018dedicated}.
      \label{fig:projections_DS}
    }
\end{figure}

Do these low-rank models suffer the same inconsistencies as the population activity model in \S\ref{sec:pop}? Here we lack a simple bound relating the mean and variance of the projections. Instead, we assess whether the model remains within the regime where mean-field theory is valid by checking the consistency of its solution. For each principal component $\boldsymbol{W}$, we solve the inverse problem and check whether the resulting mean-field parameters yield self-consistent magnetizations: 
\begin{equation} 
    \boldsymbol{\mu}_{\rm exp} \stackrel{?}{=} 
     \tanh\left[ \boldsymbol{h}_{\rm MF} + \boldsymbol{W} \ \boldsymbol{\psi}^\star\left(\boldsymbol{h}_{\rm MF}, \Lambda_{\rm MF}\right) \right] .
     \label{eq:consistency}
\end{equation} 
Compared to \S\ref{sec:pop}, an additional complexity arises from having to solve the saddle-point equations.\footnote{For repulsive patterns, corresponding to $\rho<1$, we have to remember that the solution $\psi^\star$ is complex and we have to include the two, equally important, complex conjugate saddle points. }

In real populations of neurons, we find that only components with $\rho \simeq 1$  lead to consistent solutions, but these are uninformative, corresponding to  negligible entropy reduction. More informative directions---both large and small $\rho$---are found to be inconsistent with the mean--field approximation and therefore lie outside the bounds of what a mean--field approximation can capture. As with the population activity model of \S\ref{sec:pop}, when the self-consistency condition in Eq~\eqref{eq:consistency} is violated, the effective magnetic field $\boldsymbol{h}_{\rm MF}$  tends to point systematically in the opposite direction of the magnetization. It may still be possible to fit both the mean and the variance of those principal components that violate the consistency condition, but only by introducing nearly degenerate potentials, which inherently fall outside the scope of what a mean-field approximation can capture.
These results are summarized in  Fig~\ref{fig:projections_DS}.

We could attempt to construct models using only consistent components, hoping that if we include enough of these we will make progress.  The results in \S\ref{sec:randproj}, however, show that we would need a very large number of projections to achieve a significant entropy drop, effectively defeating the purpose. This approach also reintroduces the challenge of measuring and matching a large number of observables discussed in \S\ref{sec:intro}. Moreover, similar to what occurs in Hopfield networks beyond saturation \cite{amit+al_1987}, we can expect that as the number of constrained projections increases, the model will transition away from the mean-field regime.

Alternatively, including only a few highly informative components results in models that match the large fluctuations of the experimental data by forming double-well structures in the energy landscape, as in \S\ref{sec:pop}.  Importantly, this is not a problem of the mean--field approximation but rather a feature of the data. The only way a model with the form of Eq~(\ref{eq:Eproj}) can fit the data is by being poised near a first--order phase transition. These models clearly fail to represent the data accurately.

\section{Distributional Maximum Entropy} 
\label{sec:extended}

In this section, to tame the double-well energy landscape problem we consider a maximum entropy model that matches the activity of individual neurons and the full probability distribution $P(\varphi)$ of a single projection of the population activity. We derive the mean--field equation for this new class of models and solve both the direct and the inverse problem. We apply this framework to the experimental data finding encouraging results. 

\subsection{Formulating the model}

We are interested in building a maximum entropy model that matches the 
 full probability distribution $P(\varphi)$ of activity along a single projection.  To stay in the language of Eqs~(\ref{constraint1}--\ref{maxent2}) it is useful to remember that the distribution can be written as the expectation value of a delta function,
 \begin{equation}
P(\varphi ) = {\bigg\langle} \delta \left( \varphi - \sum_{n=1}^N \frac{W_n}{\sqrt{N}} s_n\right){\bigg\rangle}.
\end{equation}
We want to match the distribution at every value of $\varphi$, so the sum over terms in Eq~(\ref{maxent2}) becomes an integral
\begin{equation}
\sum_\mu g_\mu f_\mu( \sbold ) \rightarrow N \int d\varphi U(\varphi )  \delta \left( \varphi - \sum_{n=1}^N \frac{W_n}{\sqrt{N}} s_n\right) ,
\end{equation}
where we introduce a factor of $N$ so that the ``potential'' $U(\varphi )$ is of order one.
 
The solution of the maximum entropy equations leads to the following functional form for the energy function of the model,
\begin{equation}
     E_{\rm dist}(\sbold) = - \sum_n h_n s_n + N U(\varphi) 
     \label{eq:Edist} 
\end{equation}
Here, the characteristic quadratic potential of pairwise models is replaced by a generic potential $U(\varphi)$. This potential contains in principle higher order terms $\varphi^k$ corresponding to $k$-spin interactions. In the maximum entropy construction the potential $U(\varphi)$ is fixed by the data by matching the empirical distribution  of $\varphi$, in the same way that the fields $h_n$ are fixed by matching the mean activities $\langle s_n\rangle$.  Ultimately the experimental data will tell us if these higher order interactions are relevant.  The form of the potential $U(\varphi)$ depends on the empirical probability distribution $P_{\rm exp}(\varphi)$, and some care is required in estimating this distribution from a finite data set.  In practice, for given $W_n$, we estimate $P_{\rm exp}(\varphi)$ by linearly interpolating its empirical histogram.

\subsection{Mean--field solution}

The partition function of the model is
\begin{equation} 
Z_{\rm dist} = \sum_{\sbold} \exp\left[ \sum_{n=1}^N h_n s_n - N U\left( \sum_{n=1}^N \frac{W_n}{\sqrt{N}} s_n \right) \right]  
\end{equation}
We use the  integral representation of the delta function,
\begin{equation}
\delta(x) = \int \frac {dz}{2\pi} e^{i x z},
\end{equation}
which serves to uncouple the variables $\{s_n\}$, and we find
\begin{equation} 
Z_{\rm dist} = 2^N \int \frac{dz}{2\pi} \int d\varphi \, e^{-N f_{\rm dist}(\varphi, z)} ,
\label{Zdist_int}
\end{equation}
where the local free energy is
\begin{equation} 
f_{\rm dist} (\varphi , z) = U(\varphi) + \frac{1}{N} \left[ i z \varphi - \sum_n \ln \cosh(h_n + i z \frac{W_n}{\sqrt{N}} ) \right]  . 
\label{eq:fdist}
\end{equation}
In the mean--field approximation, the integral in Eq~(\ref{Zdist_int}) is controlled by its saddle point, which obeys 
\begin{eqnarray}
0 &=& {{\partial f_{\rm dist} (\varphi , z) }\over{\partial \varphi}}{\bigg |}_{\varphi_{\rm sp}, z_{\rm sp}}\\
\Rightarrow z_{\rm sp} &=& i N U^\prime (\varphi_{\rm sp}) 
\label{zsp}
\end{eqnarray}

\begin{eqnarray}
0 &=& {{\partial f_{\rm dist} (\varphi , z) }\over{\partial z}}{\bigg |}_{\varphi_{\rm sp}, z_{\rm sp}}\\
\Rightarrow \varphi_{\rm sp} &=& \sum_{n=1}^N \frac{W_n}{\sqrt{N}} \tanh \left( h_n + i z_{\rm sp}  \right) .
\label{phisp}
\end{eqnarray}
To leading order in $1/N$ we have
\begin{equation} \ln Z_{\rm dist} = N\ln 2 - N f_{\rm dist}(\varphi_{\rm sp}, z_{\rm sp}) + \frac{1}{2} \ln \det H + \mathcal{O}
(1/N) ,
\label{eq:lnZdist}
\end{equation}
where $H$ is the Hessian, or the matrix of second derivatives of $f_{\rm dist}(\varphi,z)$ evaluated at the saddle point:
\begin{widetext}
\begin{equation} 
H = 
    \begin{pmatrix} 
        \frac {1}{N} \sum_n W_n^2 \left[ 1 - \tanh^2(h_n + i z_{\rm sp} \frac{W_n}{\sqrt{N}}) \right] & i \\
         i & N U''(\varphi_{\rm sp}) \end{pmatrix}  .
\label{eq:Hess}
\end{equation}
\end{widetext}
The Hessian does not contribute to the matching conditions at leading order, but it is important in evaluating the entropy, below.

Using this approximation we find the mean activity of each neuron 
\begin{equation}
 \langle s_n \rangle = \frac{\partial \ln Z_{\rm dist}}{\partial h_n} = \tanh\left( h_n + i z_{\rm sp} \frac{W_n}{\sqrt{N}} \right)   . \label{eq:dist_mean_sn} 
 \end{equation}
The distribution of the projected activity is defined by
\begin{equation} 
P(\varphi) = \frac{2^N}{Z_{\rm dist}} \int \frac{dz}{2\pi} e^{-N f_{\rm dist}(\varphi; z)}  . 
\end{equation}
We evaluate this in a mean--field approximation to the integral over $z$, 
\begin{equation} 
P(\varphi) = \frac{1}{Z_\varphi} \exp\left[ -N f_{\rm dist}(\varphi; z_\star (\varphi)) \right] \label{eq:Pphi} , 
 \end{equation}
which defines a $\varphi$--dependent saddle point  $z_\star(\varphi )$ as the solution of
\begin{equation} 
\varphi = \sum_{n=1}^N \frac{W_n}{\sqrt{N}} \tanh\left( h_n + i \frac{W_n}{\sqrt{N}} z_\star(\varphi) \right) .
     \label{eq:zstar} 
\end{equation}

\subsection{Inverting the mean--field equations}

The inverse problem---recovering $U(\varphi)$ and $\{h_n\}$ from data---proceeds by inverting Eqs~(\ref{eq:dist_mean_sn}) and (\ref{eq:Pphi}). This inversion is not straightforward due to the nested structure of the mean--field equations.

We begin by exploiting a gauge invariance: the energy function remains unchanged under the transformation 
\begin{eqnarray}
U(\varphi ) &\rightarrow& U(\varphi ) - \varphi U^{\prime}(\varphi_{\rm s p})\\
 h_n &\rightarrow& h_n + \sqrt{N} W_n U^{\prime}(\varphi_{\rm sp}).
\end{eqnarray}
This allows us to set $U^\prime(\varphi_{\rm sp}) = 0$, implying $z_{\rm sp} = 0$. Then Eq~(\ref{eq:dist_mean_sn}) becomes
\begin{equation} 
\langle s_n\rangle  = \tanh h_n,
 \end{equation}
as if each neuron ``felt'' the field $h_n$ with no other interactions (!).  This allows us to write the vector of fields $\boldsymbol{h} = \{h_n\}$ as
\begin{equation}
\boldsymbol{h} = {\rm atanh} (\boldsymbol{\mu}_{\rm exp}) .
\label{hdmf}
\end{equation}
It will be useful to note that in this gauge the Hessian in Eq~(\ref{eq:Hess}) becomes
\begin{equation} 
H =  \left[
\begin{array}{cc}
 \Delta & i  \\
i & N U''(\varphi_{\rm sp})      
\end{array}
 \right] ,
\label{eq:Hess3}
\end{equation}
$\Delta$ is the variance of $\varphi$  that we would find in a model of independent neurons, 
\begin{equation}
    \Delta = \frac{1}{N}\sum_n W_n^2 ( 1 - \mu_n^2) ,
    \label{chi0def}
\end{equation}
which coincides with the quantity defined in Eq~\eqref{Delta1} for a single projection. 

Next we read Equation (\ref{eq:zstar}) as an equation that allows us to numerically construct $\varphi(z_\star )$, which we  then invert to give $z_\star (\varphi )$. The result is then substituted into Eq~(\ref{eq:Pphi}) to yield an expression for the potential in the mean--field approximation,
\begin{widetext}
\begin{equation} 
         N U_{\rm MF}(\varphi) = -\ln P_{\rm exp}(\varphi)  -  \left[ i z_\star (\varphi) \varphi - \sum_n \ln \cosh\left(h_n + i z_\star (\varphi) \frac{W_n}{\sqrt{N}}\right) \right] .
\end{equation}
\end{widetext}
\subsection{Results for 1000+ neurons}

\begin{figure}
    \includegraphics[width= \linewidth]{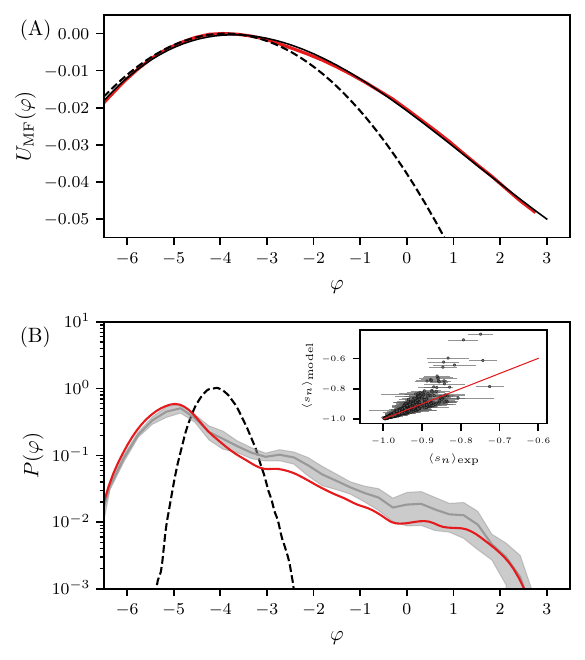}
    \caption{ 
        Maximum entropy model that matches the mean activity of each neuron and the distribution of activity along a single projection, Eq~\eqref{eq:Edist}, for $N=1416$ neurons in the mouse hippocampus. The projection corresponds to the highest variance principal component of the correlation matrix.  The experimental distribution is approximated with a $N_b = 32$ bins histogram. (A) The potential $U_{\rm MF}(\varphi )$ (solid red) compared with a quadratic (black dashed) matching the curvature of $U_{\rm MF}$ at its maximum,  and with a cubic fit (black solid).  (B) The distribution of activity $P(\varphi )$  predicted by the model (red) and estimated from the data (grey; width shows standard deviation across fifths of the data); compare with the expected results for independent neurons (dashed). Inset shows the mean activity of each neuron, model vs data. \label{fig:Uphi_Vphi}}
\end{figure}

We apply this framework to recordings from 1000+ neurons in the CA1 region of  the mouse hippocampus, as above \cite{gauthier2018dedicated,meshulam2019RG}.  In keeping with the discussion in \S\ref{sec:optimalProjections} we start by choosing $\boldsymbol{W}$ to be the principal component associated with the largest eigenvalue of the correlation matrix.  Using the mean--field approximation we infer the potential
$U_{\rm MF}(\varphi)$ shown in Fig~\ref{fig:Uphi_Vphi}A, and we notice that it is significantly different from a quadratic form.  In particular the potential is larger than quadratic at large positive $\varphi$.  While the weight vector has both positive and negative components, on average large positive $\varphi$ is associated with higher activity in the population.  Thus the non--quadratic form of the potential serves to suppress the incipient first order transition that we found in the case of models that match only the variance of activity along a single projection.

We have solved the maximum entropy problem in a mean--field approximation, and in all cases thus far this approximation has either broken down or succeeded while capturing very little of the correlation structure in the network.   To test the mean--field approximation we estimate $\{h_n\}$ and $U(\varphi )$ as above, and then do a Monte Carlo simulation of the resulting model.  If the approximation works then the mean activities $\langle s_n\rangle$ and the distribution $P(\varphi )$ that we find from this simulation should be close to what we find in the data, and this is shown in Fig \ref{fig:Uphi_Vphi}B.

We see that the agreement between theory and experiment is very good, though not perfect:  the mean--field approximation is an approximation, but a good one.  The distribution of activity along the projection is very far from what we would see if the neurons were independent.  These collective effects include a long tail toward high activity that is well described by the theory, including some structure that emerges despite the relatively featureless potential.   Importantly there is no sign of a second peak in the predicted distribution, so we have succeeded in banishing the incipient first order transition that plagued the more limited mean--field models in \S\S\ref{sec:pop} and \ref{sec:projections}.

Seeing that the mean--field approximation works, we now have to ask if these models are capturing significant structure in the patterns of network activity.  As before we use the entropy of the model, or more precisely the entropy reduction relative to a model of independent neurons, as a measure of quality (Appendices \ref{app:MaximumEntropy} and \ref{app:entropy}).

The entropy of models in the Boltzmann form of Eq~(\ref{maxent1}) is given by
\begin{equation}
S = \ln Z  + \langle E (\sbold ) \rangle . 
\end{equation}
For the distributional model defined by Eq~(\ref{eq:Edist}) we have, in the mean--field approximation,
\begin{eqnarray}
S  &=& N \ln 2 - N U_{\rm MF} (\varphi_{\rm sp}) +\sum_n \ln\cosh h_n  \nonumber\\
&&\,\,\,\,\,  +{1\over 2}\ln\det H -\sum_n h_n \langle s_n \rangle +  N\langle U_{\rm MF} (\varphi )\rangle .
\end{eqnarray}
If we do the same calculation in the independent model we have
\begin{equation}
S_0 = N \ln 2+\sum_n \ln\cosh h_n-\sum_n h_n \langle s_n \rangle ,
\end{equation}
where because of our choice of gauge the $\{h_n\}$ are the same [Eq~(\ref{hdmf})]. The entropy reduction $\Delta S = S_0 - S$ thus becomes
\begin{equation} 
    \Delta S = - N  \langle U_{\rm MF} (\varphi)  - U_{\rm MF}(\varphi_{\rm sp}) \rangle + \frac{1}{2} \ln \left[ 1 + N  U^{\prime \prime} (\varphi_{\rm sp}) \Delta \right]  . 
    \label{eq:DS_distribution}
\end{equation}
If the potential is quadratic this reduces to Eq~\eqref{eq:DS_singlePC}. In the model we are considering, where we constrain the distribution of projections along the principal component with the largest variance in activity, we find, using Eq~\eqref{eq:DS_distribution},  $\Delta S = 8.4 \pm 1.2 \, {\rm bits}$.  To set a scale, the entropy per neuron in the independent model is $S_0/N = 0.13\, {\rm bits}$. Thus by matching the distribution of just one collective coordinate we squeeze out the entropy contributed by $\sim  70$ individual neurons, or $\sim 5\%$ of the total.

\begin{figure}
    \includegraphics[width= \linewidth]{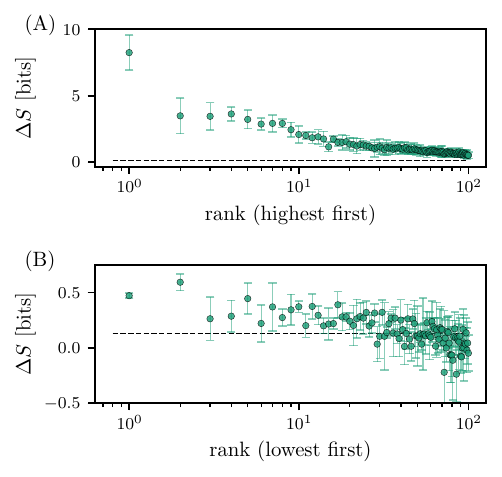}
    \caption{Entropy reductions in models  that match the distribution of individual principal components. The principal components are ranked from high to low (A) and from low to high (B).  The entropy reduction corresponding to the highest variance principal component is $\simeq 5\%$ of the entropy of the independent model. A large fraction of principal components gives an entropy reduction greater than this per-neuron independent model entropy (dashed lines).
        \label{fig:DeltaS_ranking}}
\end{figure}

We can do this calculation in models that constrain projections along the different principal components, with the results  in Fig~\ref{fig:DeltaS_ranking}.  We see that there are $\sim 100$ components that individually contribute more than $S_0/N$ to the entropy reduction, so that each of these collective coordinates is capturing more information than single neurons. Interestingly, some of the lowest variance components also yield entropy reductions $\Delta S > S_0/N$, pointing again to the relevance of repulsive patterns (\S\ref{sec:gauge1}).

\begin{figure}[b]
    \includegraphics[width = \linewidth]{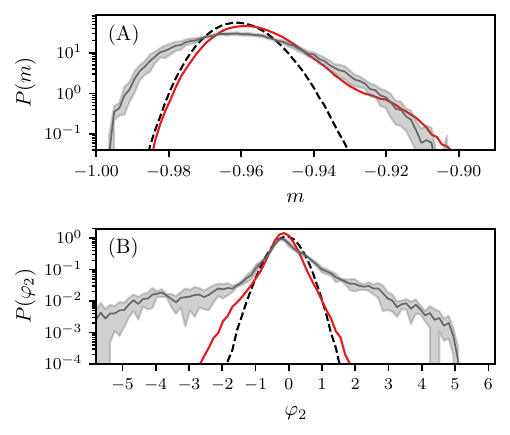}
    \caption{Testing the maximum entropy model that matches the distribution  of neural activity projected along the highest variance eigenvector of the correlation matrix. 
    (A) Distribution of the summed population activity: experimental data (gray), model prediction (red), and the independent model (dashed black). The model accurately captures the broad, non-Gaussian right tail of the distribution. 
    (B) Distribution of neural activity projected on the second principal component. The model (red) provides a slightly better fit than the independent model (dashed black), but  fails to capture much of the experimental variance---as expected, as the second principal component is, by construction, uncorrelated from the constrained projection $\varphi$.  \label{fig:exp_vs_model}}
    \end{figure}

A good model should capture statistical structure beyond the observables it explicitly constrains \cite{Meshulam+Bialek_2024}.  Here we test whether constraining the distribution of activity along the largest variance projection allows us also to predict the distribution of summed activity in the population 
\begin{equation}
m = {1\over N}\sum_{n=1}^N s_n ,
\end{equation}
or  the distribution of activity along the projection onto the second eigenvector of the covariance matrix; results are in Figs~\ref{fig:exp_vs_model}A and B, respectively.

We see in Fig~\ref{fig:exp_vs_model}A that our model does a good job of describing experimental distribution of summed population activity.  In particular  it accurately reproduces the highly non-Gaussian right hand tail, corresponding to a huge excess of high activity states relative to  what one would expect if neurons were independent.   This match extends out to states in which $\sim 4.2\%$ of neurons active simultaneously, which happens only $0.038\%$  of the time. This success is not just because the weights $W$ overlap the uniform vector, since randomizing the components of $W$ preserves this overlap but spoils the agreement. The model does less well in capturing the excess of near--silent states at the left hand tail.

In contrast to the case of the summed activity, the model does a very bad job of predicting the distribution of activity along the second principal component (Fig~\ref{fig:exp_vs_model}B).  Indeed, the predicted distribution is very similar to what we would see if the neurons were completely independent. This is perhaps not surprising, since the principal components are by definition uncorrelated (at second order), and so we expect that knowing something about one component is relatively uninformative about other components; this is not exactly true because the distributions are not Gaussian.
The relative independence of the different components suggests that we may be able to achieve near additive entropy reductions by constraining multiple projections, a point to which we will return in a subsequent paper.
 
\begin{figure}
    \includegraphics[width = \linewidth]{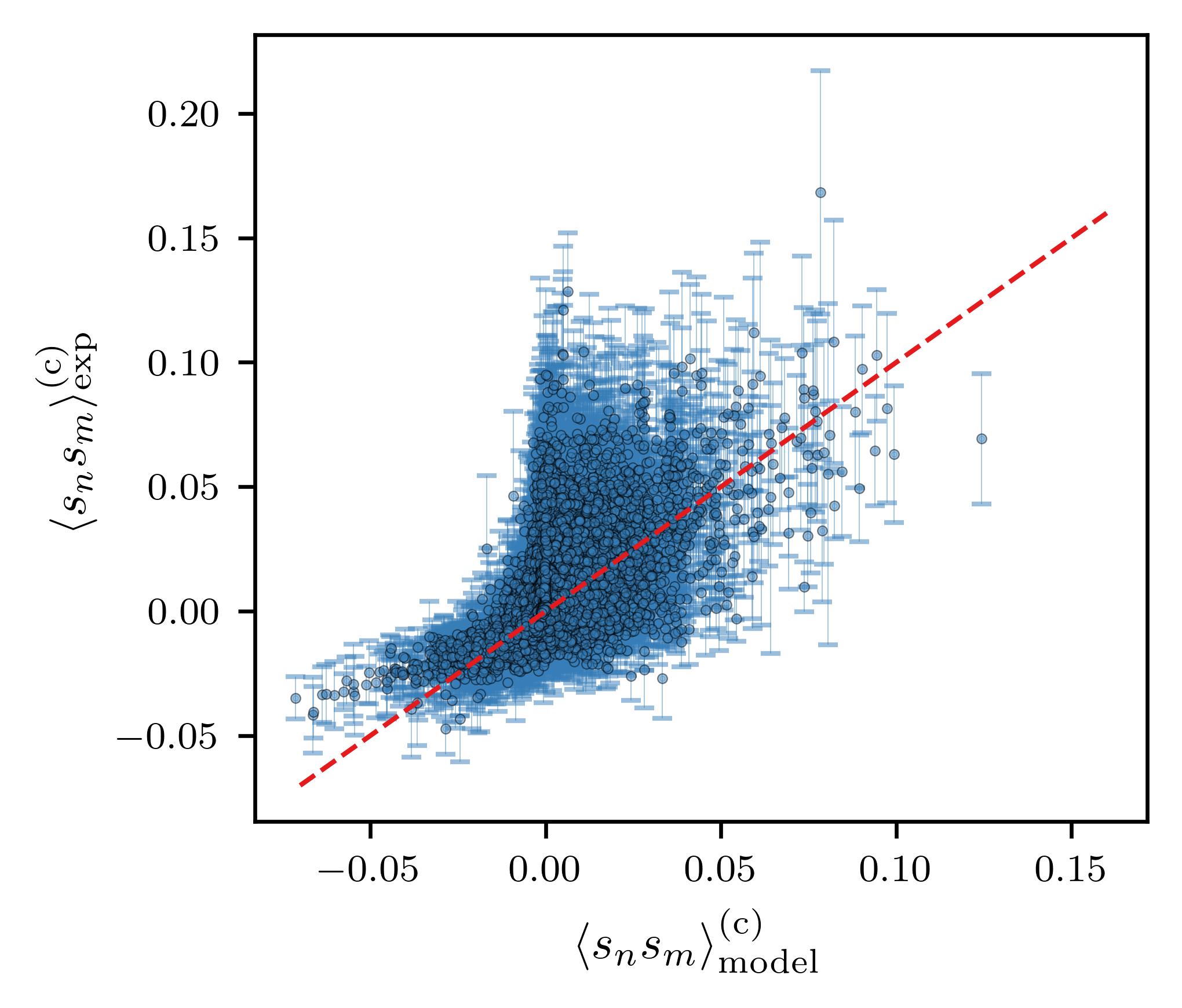}
    \caption{Connected pairwise correlations. Model vs~experimental data, with error bars from variations across fifths of the data. Although pairwise correlations were not explicitly constrained, the model successfully reproduces the overall trend and captures several of the large entries of the correlation matrix. 
        \label{fig:cij_pred}}
\end{figure}

Unlike the now conventional pairwise maximum entropy models \cite{Meshulam+Bialek_2024}, we do not match the  elements of the covariance or correlation matrix among neurons.  We do use this matrix in choosing a direction with maximum variance, and when combined with the non--quadratic form of the potential $U(\varphi )$ this makes nontrivial predictions for all $\sim N^2/2$ of the correlations despite the fact that we have only ${\cal O}(N)$ constraints.   As shown in Fig~\ref{fig:cij_pred}   the model captures the overall trend of the experimental data and reproduces several of the large entries in the correlation matrix within error bars, despite these not being explicitly constrained.   We emphasize that this is not because the covariance matrix is of low rank---the naive approximation $C \sim \Delta WW^T$ fails completely.  We conclude that the non-quadratic terms in $U(\phi )$ are making it possible for a model that focuses on a single projection to make rough but non-trivial predictions beyond the rank one approximation to the covariance.

Finally we can ask where these statistical physics models for neural activity sit in the phase diagram of possible models with the same general form.  In \S\S\ref{sec:pop} and \ref{sec:projections} we saw that trying to match measured expectation values drove simpler mean--field models toward a first order phase transition, which is interesting but in qualitative disagreement with other features of the data.  The distributional maximum entropy models that we find are far from any first order transitions, but touch a (second order) critical point at parameters where the determinant of the Hessian in Eq~(\ref{eq:Hess3}) vanishes.  This condition is
\begin{equation}
1+ N U''(\varphi_{\rm sp}) \Delta= 0 ,
\label{critline}
\end{equation}
where again $\Delta$ is the variance of $\varphi$ that we expect from independent neurons, as in Eq~(\ref{chi0def}).  

\begin{figure}[b]
    \includegraphics[width=\linewidth]{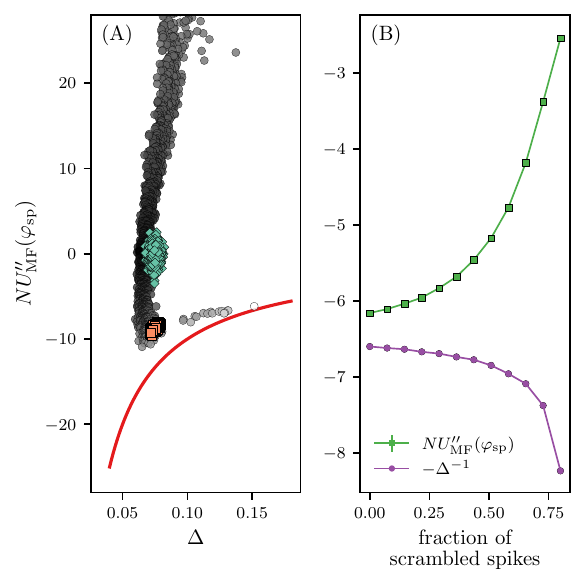}
    \caption{Approach to a critical point in matching the distribution of activity along a projection,  for $N=1416$ neurons in the mouse hippocampus.   (A) Scatter plot of $NU^{\prime \prime} _{\rm MF} (\varphi_{\rm sp})$ vs $\Delta$ for different projections: random (green), random positive (red), and the eigenvectors (Grey scale: lighter shades indicate higher $\Delta S$) of the correlation matrix.  The red line marks the critical parameter settings where $1 + NU''(\varphi_{\rm sp}) \Delta  = 0$.
    (B) Trajectories of  $N U^{\prime \prime} _{\rm MF} (\varphi_{\rm sp})$  and $-\Delta ^{-1}$ as we look at networks with same mean activity for each neuron but weaker correlations, generated by shuffling a fraction of the spikes independently for each neuron. 
       \label{fig:critical}}
\end{figure}

We have emphasized a model that matches the distribution of activity along the dominant principal component, but it is useful to ask what happens if we redo the analysis with different choices for this projection.  As shown in Fig~\ref{fig:critical}A, for random choices of the projection weights $W_n$ the model is far from criticality.  If we bias the weights to all be positive, we get closer to criticality but still some distance away. If we choose the $W_n$ to be the eigenvectors of the covariance matrix, then as we look at components that generate larger and larger reductions of the entropy (Fig~\ref{fig:DeltaS_ranking}) we see a sequence of modes that approaches the critical line defined by Eq~(\ref{critline}). For the dominant mode we have
\begin{equation}
\min \left[1+  N U''(\varphi_{\rm sp}) \Delta\right] = 0.06,
\end{equation}
so that matching expectation values drives the model to within a few percent of the critical point.

The approach to criticality depends on the strength of correlations in the network.  We can imagine systems in which the mean activity of each neuron is the same, but the correlations between pairs of neurons are  weaker, and we can generate   such data by   shuffling a fraction of the time bins independently for each neuron.  For each shuffled data set we repeat the construction above,  and we find that N$U^{\prime \prime} _{\rm MF} (\varphi_{\rm sp})$ and $-\chi_0^{-1}$ gradually move apart as we consider less correlated networks.  Strikingly, a $\sim 20\%$ reduction in the strength of correlations pushes the model a factor of two further away from criticality: plausible populations of neurons would be farther from criticality than the real network. This is consistent with other signatures of near--critical behavior \cite{Meshulam+Bialek_2024}, including the scaling of these same data under coarse--graining \cite{meshulam2019RG}.

\section{Conclusion}

If we think that the dynamics of large neural populations are dominated by a small number of collective variables, it is tempting to write mean--field models for the distribution over network states.  Beginning with the simplest example---a model constraining only the mean and variance of the summed populations activity---we demonstrated that experimental neural populations systematically sit outside of the bound of what can be described by the mean--field approximation. Matching the empirical moments drives the parameters of these models close to a first--order phase transition, characterized by a double-well structure in its energy landscape.  This structure leads to qualitative disagreements with the observed distribution of activity. 

We then extended this approach to models constraining the mean activity of each individual neuron and the variance along multiple projections of the neural activity. Choosing these projections at random yields mathematically consistent, yet empirically uninformative, models. Optimal projection selection, guided by the miniMax entropy principle, partially resolves this issue but reveals that informative directions inevitably lead to double-well energy landscapes, reflecting the same limitations seen in the population activity model.

To address these fundamental shortcomings, we proposed a new class of distributional maximum entropy models, constraining not just means or variances but the full empirical distributions of neural activity along  projections. Matching the empirical distribution of these projections requires fitting a potential which contains higher--order interactions, moving beyond the pairwise quadratic assumptions inherent in traditional mean-field approaches. We successfully applied this model to experimental data from the mouse hippocampus.  The mean--field approximation is (finally) internally consistent and the resulting model captures strong correlation structures in the data.    Furthermore, the model predicts features of the data that were not used in its construction, such as the distribution of the population activity and individually strong pairwise correlations .

Finally, our analysis predicts that many principal components are, in principle, highly informative. This suggests that extending the distributional framework to multiple projections may yield even more powerful models, paving the way for scalable and accurate analysis of large-scale neural population recordings.       
\begin{acknowledgments}
We thank our experimentalist colleagues MJ Berry II, CD Brody, JL Gauthier, O Marre, and DW Tank for  guiding us through  the data.  LDC thanks FG Castro for useful comments on the manuscript. Work  supported in part by the  National Science Foundation, through the Center for the Physics of Biological Function (PHY--1734030); and by fellowships from the Human Frontiers Science Program (FM),   the James S.~McDonnell Foundation (CWL), the John Simon Guggenheim Memorial Foundation (WB),  and the Simons Foundation (WB and FM).
\end{acknowledgments}

\appendix

\section{Entropy, likelihood, and model quality}
\label{app:MaximumEntropy}

Here we collect some results on maximum entropy, maximum likelihood, measures of model quality, and the miniMax entropy principle.  None of these results are new, but since they form essential background we thought it would be useful to collect them here in  language as close as possible to that in the main text.  Some of these ideas also appear in a recent review \cite{carcamo+al_2025}.

How do we measure the performance of a model in describing data?  One simple idea is the measure the probability that the model generates the data we have observed.  In our case the states of the network are defined by $\sbold$, and if we observe a set of $N_s$ independent samples 
\begin{equation}
\{\sbold^{(i)}\} \equiv \{\sbold^{(1)},\,\sbold^{(2)},\cdots ,\,\sbold^{(N_s)}\} 
\end{equation}
then the log probability of the data in a model $P(\sbold )$ is the normalized likelihood
\begin{equation}
{\cal L} = {1\over {N_s}}\sum_{i=1}^{N_s} \ln P(\sbold^{(i)} )  = \langle  \ln P(\sbold^{(i)} )\rangle_{\rm exp} .
\end{equation}
The maximum likelihood principle is that we should choose the model, and its parameters, that maximizes $\cal L$ \cite{bialek2012biophysics}.

Consider the class of maximum entropy models that can match expectation values of observables $\{f_\mu (\sbold )\}$, as in Eqs~(\ref{maxent1}, \ref{maxent2}).  Then if we evaluate the likelihood of the data in this model we find
\begin{eqnarray}
{\cal L} &\equiv&  {1\over {N_s}}\sum_{i=1}^{N_s} \ln P(\sbold^{(i)} )\nonumber\\
&=&  {1\over {N_s}}\sum_{i=1}^{N_s} \ln\left(  \frac 1 Z \exp\left[ -E(\sbold^{(i)}) \right]  \right)\\
&=& -\ln Z  - {1\over {N_s}}\sum_{i=1}^{N_s}E(\sbold^{(i)})\\
&=& - \ln Z  - \sum_\mu g_\mu \langle f_\mu (\sbold )\rangle_{\rm exp} .
\end{eqnarray}
If we ask for the values of the couplings $\{g_\mu\}$ that maximize the likelihood we should solve the equations
\begin{eqnarray}
{{\partial{\cal L}}\over{\partial g_\mu}} &=& 0\\
\Rightarrow - {{\partial{\ln Z}}\over{\partial g_\mu}} &=& \langle f_\mu (\sbold )\rangle_{\rm exp} .
\end{eqnarray}
But for models with the Boltzmann form in Eq~(\ref{maxent1}) we have thermodynamic identities
\begin{equation}
- {{\partial{\ln Z}}\over{\partial g_\mu}} = \langle f_\mu (\sbold )\rangle_{P} .
\end{equation}
Thus if we view the {\em form} of the maximum entropy model as given, adjusting the parameters to maximize the likelihood is the same as imposing the constraints
\begin{equation}
\langle f_\mu (\sbold )\rangle_{P} = \langle f_\mu (\sbold )\rangle_{\rm exp}
\end{equation}

If we have a probability distribution $P(\sbold )$ then we can construct a code in which each state $\sbold$ is represented by a codeword of length $\ell (\sbold ) = -\ln P(\sbold )$ \cite{shannon1948mathematical,cover+thomas_91}.  The model $P(\sbold )$ thus allows us to describe the data with an average code length per sample
\begin{equation}
\bar \ell = - {1\over {N_s}}\sum_{i=1}^{N_s} \ln P(\sbold^{(i)} )  = {\bigg\langle}  \left[ - \ln P(\sbold^{(i)} )\right] {\bigg\rangle}_{\rm exp} .
\end{equation}
Another natural principle is that we prefer models that give the greatest compression of the data, or the shortest description.  We see that minimizing this code length is the same as maximizing the likelihood.

We can use the equivalence of code length and (negative) likelihood to write
\begin{eqnarray}
\bar \ell &=& -{\cal L} = \ln Z  + \sum_\mu g_\mu \langle f_\mu (\sbold )\rangle_{\rm exp} \\
&=& \ln Z  + \sum_\mu g_\mu \langle f_\mu (\sbold )\rangle_{P}\\
&=& \ln Z + \langle E(\sbold )\rangle_P .
\end{eqnarray}
Further, the partition function is related to the free energy, $F = -\ln Z$, and since Eq~(\ref{maxent1}) is a Boltzmann distribution in which the  temperature $k_BT = 1$, we have 
\begin{equation}
F = \langle E(\sbold )\rangle_P - S[P(\sbold )].
\end{equation}
Putting these together we find
\begin{equation}
\bar \ell = S[P(\sbold  )].
\end{equation}
Thus for maximum entropy distributions  (though not in general!) the mean code length evaluated on the data is the entropy of our model.

We still have the principle of minimizing the code length. If we can choose different constraints, we see that this can be accomplished by choosing the ones for which the maximum entropy has the minimum value: the miniMax entropy principle.  This idea has roots in work on computational vision from the 1990s \cite{zhu+al_1997} but seems not to have been widely appreciated.  In general, implementing the miniMax entropy principle is challenging, and one can make progress only by searching over limited classes of constraints, as with the different projections considered here or tree--like patterns of connectivity \cite{Lynn2023,Lynn+al_2025b}.

\section{Corrections to mean--field}
\label{app:corrections}

Corrections to   the mean--field approximation arise from integrating over the fluctuations around the saddle point \cite{bender2013advanced}.  The leading term, as seen in  Eq~\eqref{eq:projection_freeEnergy}, is logarithm that comes from a Gaussian approximation to the integral, and corresponds to one--loop diagrams in field theory.
Independent of these approximations, the mean activity of each neuron is given by 
\begin{eqnarray} 
    \mean{s_n} = - \frac{\partial F ( \bh, \Lambda) }{\partial h_n } 
\end{eqnarray}
When differentiating the one--loop free energy in Eq~\eqref{eq:projection_freeEnergy} with respect to $h_n$ we have to carefully consider all the $h_n$ dependencies. In fact the term $\Delta$ in the logarithm depends explicitly on $h_n$, so that 
\begin{equation}
    -\frac { \partial F} { \partial h_n} = \bar \mu_n - \frac 1 N \sum_{m} R_{nm} \bar \mu_m  \frac{ \partial \bar \mu_m}{ \partial  h_n}  \, 
\end{equation}
where the matrix $R$ is defined as
\begin{equation}
     R =  W^T ( \mathbb I - \Lambda \Delta ) ^{-1} \Lambda W
\end{equation}
where, to lighten the notation, we are using $\bar \mu_n$ instead of $\mu_n ^{(0)}$ to denote the leading term in the magnetization. 

The derivative of the zeroth order magnetization $\bar \mu_m$ with respect to the external field is given by, 
\begin{eqnarray}
\frac{ \partial \bar \mu_m}{      \partial h_n} =  \sum_{m, \gamma}  \left[ \delta_{nm} + W_{\gamma m} \frac{ \partial \psi^\star_{\gamma}}{\partial h_n} \right] \left( 1-\bar \mu_m^2 \right) .
\label{app:eqdmudh}
\end{eqnarray}
This requires computing the derivative of the fixed point $\psi^\star$ with respect to the external fields $h_n$,
\begin{equation}
    K_{\alpha  n} = \frac{ \partial \psi_{\alpha} ^\star(\bh, \Lambda) }{ \partial h_n }  .
\end{equation}
We can compute this  by differentiating the saddle point Eq~\eqref{eq:projection_SP} and rearranging, to find 
\begin{equation}
 K_{\alpha n}  = \frac 1 N \sum_{\beta, \gamma, m} \Lambda_{\alpha \beta} W_{\beta m}\left( \delta_{nm} +  W_{\gamma m}  K_{\gamma n} \right) \left[ 1 - \bar \mu_m ^2 \right] . 
\end{equation}
  Using the definition of $\Delta$ from Eq~\eqref{Delta1} this becomes
\begin{eqnarray}
    K = \Lambda \Delta K + \frac 1 N \Lambda W \operatorname{diag}\left[ 1 - \bar{\boldsymbol{\mu}} ^2  \right] ,
\end{eqnarray}
which leads to 
\begin{eqnarray}
    K = \frac {1} {N} \left( \mathbb I - \Lambda \Delta\right)^{-1}  \Lambda W \operatorname{diag}\left[ 1 - \bar{\boldsymbol{\mu}}^2 \right] .
\end{eqnarray}
Substituting this into Eq~\eqref{app:eqdmudh} we find 
\begin{eqnarray}
    \frac{\partial \bar \mu_m}{\partial h_n} = \sum_m \left[ \delta_{nm} + \frac 1 N R_{nm} \left( 1 - \bar \mu_m^2 \right)\right] \left( 1 - \bar \mu_m^2 \right) .
\end{eqnarray}
Putting the pieces together we have
\begin{eqnarray}
    \mu^{(1)}_n &=& \bar \mu_n - \frac 1 N R_{nn} \bar \mu_n ( 1- \bar \mu _n^2)  \nonumber\\
    &&\,\,\,\,\,\,\,\,\,- \frac 1 {N^2} \sum_{m = 1}^{N} R_{nm}^2 \bar \mu_m \left( 1- \bar \mu_m^2\right) ^2 ,
\end{eqnarray}
 where, since the elements of the matrix $R_{nm}$ are $\mathcal{O}\left(1\right)$, the last term on the right-hand side is $\mathcal{O}\left(1/N\right)$.

\section{Computation of the entropy} 
\label{app:entropy}

Here we collect results on the entropy of various models.

\subsection{Independent model}
The maximum-entropy model assuming independent neurons and only matching the mean activities $\mu_n=\langle s_n\rangle$ has energy
\begin{align}
    E_{0}(\mathbf{s})=-\sum_{n=1}^Nh_ns_n\;.
\end{align}
The partition function can be computed exactly 
\begin{align}
 Z_{0}=\prod_{n=1}^N\sum_{s_n=\pm 1} \exp\left(h_n s_n\right)=2^N\prod_{n=1}^N\cosh(h_n)\;.
\end{align}
Therefore, the fields are given by 
\begin{align}
    \mu_n=\frac{{\rm d}\ln Z_0}{{\rm d}h_n}=\tanh(h_n) \Longrightarrow h_n = {\rm atanh}(\mu_n),
\end{align}
and the entropy is 
\begin{eqnarray}
   S_{0}&=&\ln Z_0 + \langle E_0(\mathbf{s})\rangle\\ 
   &=&N\ln 2 -\sum_{n=1}^N h_n\mu_n \nonumber\\
   &&\,\,\,\,\,\,\,\,\,  + \sum_{n=1}^N \ln\cosh{\rm atanh}(\mu_n) .
\label{eq:S0_appendix}
\end{eqnarray}

\subsection{Pairwise projection model} 
The entropy of the pairwise model in Eq~\eqref{eq:Eproj} is  
\begin{widetext}
\begin{equation}
\label{eq:Sproj_appendix}
    S_{\rm proj}=\ln Z_{\rm proj} + \langle E_{\rm proj}(\mathbf{s})\rangle\\
    = \ln Z_{\rm proj} -\sum_{n=1}^Nh_n\mu_n \\-\frac{1}{2N}\sum_{\alpha,\beta=1}^K\Lambda_{\alpha\beta}\left(\chi_{\alpha\beta}+\sum_{n,m=1}^N W_{\alpha n}\mu_nW_{\beta m}\mu_m\right),
\end{equation}
where we have used the definition in Eq~\eqref{eq:Eproj} and the maximum entropy property. We can use the saddle point approximation of the free energy from Eq~\eqref{eq:projection_freeEnergy} to write 
\begin{eqnarray}
    \ln Z_{\rm proj} &\simeq& -N f_{\rm proj}(\boldsymbol{\psi}^\star) - \frac{1}{2} \ln \left| \mathbb{I} - \Lambda \Delta \right|\\
    &=& N\ln 2-\frac 1 {2N}\sum_{\alpha,\beta=1}^K\sum_{n,m=1}^N\Lambda_{\alpha\beta}W_{\alpha n}\mu_nW_{\beta m}\mu_m 
   +\sum_{n=1}^N\ln\cosh{\rm atanh}(\mu_n) -\frac 12 \ln \left|\chi^{-1}\Delta\right|,\label{eq:lnZproj_appendix}
\end{eqnarray}
\end{widetext}
where in the second equality we have used the expressions for the couplings and fields from Eqs~(\ref{eq:projections_Lambda},\ref{eq:projections_h}), obtained in the mean-field approximation. Therefore, substituting Eq~\eqref{eq:lnZproj_appendix} into Eq~\eqref{eq:Sproj_appendix} and using Eq~\eqref{eq:projections_h} together with the expression for the entropy of the independent model in Eq~\eqref{eq:S0_appendix}, to leading order in $1/N$, we find 
\begin{align}
    \begin{split}
        S_{\rm proj}\simeq S_0 -\frac 12 {\rm Tr}[\Delta^{-1}\chi-\ln\left(\Delta^{-1}\chi\right)-\mathbb{I}] \,,\label{eq:Sproj_appendix_final}
    \end{split}
\end{align}
recovering the expression for $\Delta S(W)=S_0-S$ in Eq~\eqref{eq:proj_DS}.
\subsection{Distributional projection model}
The entropy for the distributional model in Eq~\eqref{eq:Edist} is 
\begin{eqnarray}
    S_{\rm dist} & =& \ln Z_{\rm dist} +\langle E_{\rm dist}(\mathbf{s})\rangle\\
  &  =&\ln Z_{\rm dist} -\sum_nh_n \mu_n+N \langle U (\varphi)\rangle .
  \end{eqnarray}
Using the mean-field approximation to leading order in $1/N$ from Eq~\eqref{eq:lnZdist}, we have
\begin{eqnarray}
S_{\rm dist}    &\simeq& N\ln 2 -Nf_{\rm dist}(\varphi_{\rm sp},z_{\rm sp})-\frac 12 \ln\det H \nonumber \\
    &&\,\,\,\,\,\,\,\,\,  -\sum_nh_n \mu_n+N \langle U (\varphi)\rangle\,,
\end{eqnarray}
where the Hessian $H$ is defined in Eq~\eqref{eq:Hess}.  Note that, in this case, the mean-field solution for the fields is $h_n={\rm atanh}(\mu_n)$, as derived in Eq~\eqref{hdmf}. If we isolate the terms corresponding to the entropy of the independent model in Eq~\eqref{eq:S0_appendix}, we find
\begin{eqnarray}
    S_{\rm dist} &\simeq & S_0+N\left(\langle U (\varphi)\rangle-U(\varphi_{\rm sp})\right)\nonumber\\
   &&\,\,\,\,\,\,\,\,\,\,  -  \frac 12 \ln\left(1+NU''(\varphi_{\rm sp})\Delta \right)\,.
\end{eqnarray}
If we expand the potential up to second order, and recall that we have chosen a gauge where $U'(\varphi_{\rm sp})=0$, we obtain
\begin{eqnarray}
    S_{\rm ind} &\simeq& S_0 +\frac 12 \frac{N U''(\varphi_{\rm sp})\Delta}{1+N U''(\varphi_{\rm sp})\Delta}\nonumber\\
    &&\,\,\,\,\,\,\,\,\, -\frac 12 \ln\left(1+NU''(\varphi_{\rm sp})\Delta \right) .
    \label{eq:Sdist_appendix_expansion}
\end{eqnarray}
Finally, we can compute the fluctuations $\chi$ from the inverse Hessian in Eq~\eqref{eq:Hess}:
\begin{align}
    \chi=\frac{\Delta}{1+NU''(\varphi_{\rm sp})\Delta}\,.
\end{align}
Therefore, Eq~\eqref{eq:Sdist_appendix_expansion} is equivalent to Eq~\eqref{eq:Sproj_appendix_final} obtained for the pairwise potential.

\section{Entropy reduction with a single random projection}
\label{app:rand}

Here we show that the entropy reduction of the pairwise model with a single random projection vanishes as the system size tends to infinity, as discussed at the end of \S\ref{sec:randproj}. We consider a single random projection with i.i.d.\ Gaussian elements of the weight vector $W_n\sim\mathcal{N}(0,1)$. The entropy reduction, from Eq~\eqref{eq:proj_DS} of the main text, is 
\begin{widetext}
\begin{equation}
    \Delta S (W)= \frac 12 \left[\frac 1 N \sum_{n,m=1}^N W_n \tilde C_{nm} W_m - \ln\left(\frac 1 N \sum_{n,m=1}^N W_n \tilde C_{nm} W_m\right)  -1\right].
\end{equation}
\end{widetext}
We want to show that the entropy reduction is zero in the limit of large $N$, which is equivalent to the statement that
\begin{equation}
\underset{N\rightarrow\infty}   \lim \frac 1 N \sum_{n,m}W_n \tilde C_{nm} W_m=1 .\label{eq:limit}
\end{equation}
To this end, we decompose the vector $W$ in the basis of the eigenvectors $u_{\alpha n}$ ($\alpha,n=1,\ldots,N$) of the correlation matrix $\tilde C$, with eigenvalues $\rho_\alpha$:
\begin{align}
    W_n=\sum_{\alpha=1}^N\lambda_\alpha\,u_{\alpha n}\;, &&\text{where}&& \lambda_\alpha=\sum_{n=1}^N u_{\alpha n}W_n\;.
\end{align}
Taking expectations over $W$, we find
\begin{eqnarray}
\mathbb{E}_W\left[\lambda_\alpha\right]&=&0 , \\
\mathbb{E}_W\left[\lambda_\alpha^2\right]&=& u_\alpha^\top \mathbb{E}_W\left[W W^\top\right]u_\alpha=1 ,
\end{eqnarray}
where we have used that the eigenvectors $u_\alpha$ are orthonormal. 

To make progress toward Eq~(\ref{eq:limit}) we first write
\begin{align}
   \frac 1N W^\top \tilde C W=\frac 1 N \sum_{\alpha,\beta=1}^N\lambda_\alpha\lambda_\beta\rho_\alpha\delta_{\alpha,\beta}=\frac 1 N\sum_{\alpha=1}^N\lambda_\alpha^2 \rho_\alpha\;.
\end{align}
The first moment is then
\begin{equation}
\frac 1 N \sum_{\alpha=1}^N\mathbb{E}_W\left[\lambda_\alpha^2 \right]\rho_\alpha=\frac 1 N \sum_{\alpha=1}^N \rho_\alpha =1 ,
\end{equation}
where we have used that ${\rm Tr}[\tilde C]=\sum_{\alpha=1}^N\rho_\alpha=N$. This shows that Eq~(\ref{eq:limit}) is true on average.

To check that fluctuations don't spoil the result, we look at the second moment:
\begin{widetext}
    \begin{equation}
    \frac 1 {N^2}\mathbb{E}_W\left[\left(W^\top \tilde C W\right)^2\right] = \frac 1 {N^2}\sum_{\alpha\beta}\rho_\alpha\rho_\beta\mathbb{E}_W\left[\lambda_\alpha^2\lambda_\beta^2\right]
    =
    \frac 1 {N^2} \sum_{\alpha\beta} \rho_\alpha\rho_\beta \sum_{ijkl} u_{\alpha i} u_{\alpha j} u_{\beta k} u_{\beta l}\mathbb{E}_W\left[W_iW_jW_kW_l\right] .
     \end{equation}
    \end{widetext}
     Using
 \begin{equation}
\mathbb{E}_W\left[W_iW_jW_kW_l\right] = \left(\delta_{ij} \delta_{kl} + \delta_{ik} \delta_{il} + \delta_{il} \delta_{jk} \right)
\end{equation}
we obtain
\begin{equation}
\frac 1 {N^2}\mathbb{E}_W\left[\left(W^\top \tilde C W\right)^2\right] =  1+\frac 2N .
    \end{equation}
Thus the variance of fluctuations around the equality in Eq~(\ref{eq:limit}) are vanishing as $\sim 1/N$.

\bibliography{refs_MFT}

\begin{thebibliography}{46}
\expandafter\ifx\csname natexlab\endcsname\relax\def\natexlab#1{#1}\fi
\expandafter\ifx\csname bibnamefont\endcsname\relax
  \def\bibnamefont#1{#1}\fi
\expandafter\ifx\csname bibfnamefont\endcsname\relax
  \def\bibfnamefont#1{#1}\fi
\expandafter\ifx\csname citenamefont\endcsname\relax
  \def\citenamefont#1{#1}\fi
\expandafter\ifx\csname url\endcsname\relax
  \def\url#1{\texttt{#1}}\fi
\expandafter\ifx\csname urlprefix\endcsname\relax\def\urlprefix{URL }\fi
\providecommand{\bibinfo}[2]{#2}
\providecommand{\eprint}[2][]{\url{#2}}

\bibitem[{\citenamefont{{Allen Institute MindScope Program Allen Brain
  Observatory}}()}]{AllenData}
\bibinfo{author}{\bibnamefont{{Allen Institute MindScope Program Allen Brain
  Observatory}}}, \emph{\bibinfo{title}{Neuropixels visual coding (dataset)
  2019}}, \bibinfo{howpublished}{https://brain-map.org/explore/circuits}.

\bibitem[{\citenamefont{Gauthier and Tank}(2018)}]{gauthier2018dedicated}
\bibinfo{author}{\bibfnamefont{J.~L.} \bibnamefont{Gauthier}} \bibnamefont{and}
  \bibinfo{author}{\bibfnamefont{D.~W.} \bibnamefont{Tank}},
  \bibinfo{journal}{Neuron} \textbf{\bibinfo{volume}{99}}, \bibinfo{pages}{179}
  (\bibinfo{year}{2018}).

\bibitem[{\citenamefont{Manley et~al.}(2024)\citenamefont{Manley, Lu, Barber,
  Demas, Kim, Meyer, Traub, and Vaziri}}]{manley2024simultaneous}
\bibinfo{author}{\bibfnamefont{J.}~\bibnamefont{Manley}},
  \bibinfo{author}{\bibfnamefont{S.}~\bibnamefont{Lu}},
  \bibinfo{author}{\bibfnamefont{K.}~\bibnamefont{Barber}},
  \bibinfo{author}{\bibfnamefont{J.}~\bibnamefont{Demas}},
  \bibinfo{author}{\bibfnamefont{H.}~\bibnamefont{Kim}},
  \bibinfo{author}{\bibfnamefont{D.}~\bibnamefont{Meyer}},
  \bibinfo{author}{\bibfnamefont{F.~M.} \bibnamefont{Traub}}, \bibnamefont{and}
  \bibinfo{author}{\bibfnamefont{A.}~\bibnamefont{Vaziri}},
  \bibinfo{journal}{Neuron} \textbf{\bibinfo{volume}{112}},
  \bibinfo{pages}{1694} (\bibinfo{year}{2024}).

\bibitem[{\citenamefont{Meshulam and Bialek}(2024)}]{Meshulam+Bialek_2024}
\bibinfo{author}{\bibfnamefont{L.}~\bibnamefont{Meshulam}} \bibnamefont{and}
  \bibinfo{author}{\bibfnamefont{W.}~\bibnamefont{Bialek}},
  \bibinfo{journal}{arXiv preprint arXiv:2409.00412}  (\bibinfo{year}{2024}).

\bibitem[{\citenamefont{Jaynes}(1957)}]{jaynes1957information}
\bibinfo{author}{\bibfnamefont{E.~T.} \bibnamefont{Jaynes}},
  \bibinfo{journal}{Physical Review} \textbf{\bibinfo{volume}{106}},
  \bibinfo{pages}{620} (\bibinfo{year}{1957}).

\bibitem[{\citenamefont{Jaynes}(1982)}]{jaynes1982rationale}
\bibinfo{author}{\bibfnamefont{E.~T.} \bibnamefont{Jaynes}},
  \bibinfo{journal}{Proceedings of the IEEE} \textbf{\bibinfo{volume}{70}},
  \bibinfo{pages}{939} (\bibinfo{year}{1982}).

\bibitem[{\citenamefont{M\'ezard et~al.}(1987)\citenamefont{M\'ezard, Parisi,
  and Virasoro}}]{mezard+al_87}
\bibinfo{author}{\bibfnamefont{M.}~\bibnamefont{M\'ezard}},
  \bibinfo{author}{\bibfnamefont{G.}~\bibnamefont{Parisi}}, \bibnamefont{and}
  \bibinfo{author}{\bibfnamefont{M.~A.} \bibnamefont{Virasoro}},
  \emph{\bibinfo{title}{Spin Glass Theory and Beyond}}
  (\bibinfo{publisher}{World Scientific, Singapore}, \bibinfo{year}{1987}).

\bibitem[{\citenamefont{Ackley et~al.}(1985)\citenamefont{Ackley, Hinton, and
  Sejnowski}}]{ackley+al_85}
\bibinfo{author}{\bibfnamefont{D.~H.} \bibnamefont{Ackley}},
  \bibinfo{author}{\bibfnamefont{G.~E.} \bibnamefont{Hinton}},
  \bibnamefont{and} \bibinfo{author}{\bibfnamefont{T.~J.}
  \bibnamefont{Sejnowski}}, \bibinfo{journal}{Cognitive Science}
  \textbf{\bibinfo{volume}{9}}, \bibinfo{pages}{147} (\bibinfo{year}{1985}).

\bibitem[{\citenamefont{Hopfield}(1982)}]{hopfield1982}
\bibinfo{author}{\bibfnamefont{J.~J.} \bibnamefont{Hopfield}},
  \bibinfo{journal}{Proceedings of the National Academy of Sciences (USA)}
  \textbf{\bibinfo{volume}{79}}, \bibinfo{pages}{2554} (\bibinfo{year}{1982}).

\bibitem[{\citenamefont{Hopfield and Tank}(1985)}]{hopfield+tank1985}
\bibinfo{author}{\bibfnamefont{J.~J.} \bibnamefont{Hopfield}} \bibnamefont{and}
  \bibinfo{author}{\bibfnamefont{D.~W.} \bibnamefont{Tank}},
  \bibinfo{journal}{Biological Cybernetics} \textbf{\bibinfo{volume}{52}},
  \bibinfo{pages}{141} (\bibinfo{year}{1985}).

\bibitem[{\citenamefont{Hopfield and Tank}(1986)}]{hopfield+tank1986}
\bibinfo{author}{\bibfnamefont{J.~J.} \bibnamefont{Hopfield}} \bibnamefont{and}
  \bibinfo{author}{\bibfnamefont{D.~W.} \bibnamefont{Tank}},
  \bibinfo{journal}{Science} \textbf{\bibinfo{volume}{233}},
  \bibinfo{pages}{625} (\bibinfo{year}{1986}).

\bibitem[{\citenamefont{Amit et~al.}(1987)\citenamefont{Amit, Gutfreund, and
  Sompolinsky}}]{amit+al_1987}
\bibinfo{author}{\bibfnamefont{D.~J.} \bibnamefont{Amit}},
  \bibinfo{author}{\bibfnamefont{H.}~\bibnamefont{Gutfreund}},
  \bibnamefont{and}
  \bibinfo{author}{\bibfnamefont{H.}~\bibnamefont{Sompolinsky}},
  \bibinfo{journal}{Annals of Physics} \textbf{\bibinfo{volume}{173}},
  \bibinfo{pages}{30} (\bibinfo{year}{1987}).

\bibitem[{\citenamefont{Histed and Maunsell}(2014)}]{histed2014cortical}
\bibinfo{author}{\bibfnamefont{M.~H.} \bibnamefont{Histed}} \bibnamefont{and}
  \bibinfo{author}{\bibfnamefont{J.~H.} \bibnamefont{Maunsell}},
  \bibinfo{journal}{Proceedings of the National Academy of Sciences}
  \textbf{\bibinfo{volume}{111}}, \bibinfo{pages}{E178} (\bibinfo{year}{2014}).

\bibitem[{\citenamefont{Rule et~al.}(2019)\citenamefont{Rule, O’Leary, and
  Harvey}}]{rule2019causes}
\bibinfo{author}{\bibfnamefont{M.~E.} \bibnamefont{Rule}},
  \bibinfo{author}{\bibfnamefont{T.}~\bibnamefont{O’Leary}},
  \bibnamefont{and} \bibinfo{author}{\bibfnamefont{C.~D.}
  \bibnamefont{Harvey}}, \bibinfo{journal}{Current Opinion in Neurobiology}
  \textbf{\bibinfo{volume}{58}}, \bibinfo{pages}{141} (\bibinfo{year}{2019}).

\bibitem[{\citenamefont{Bialek et~al.}(2012)\citenamefont{Bialek, Cavagna,
  Giardina, Mora, Silvestri, Viale, and Walczak}}]{bialek2012statistical}
\bibinfo{author}{\bibfnamefont{W.}~\bibnamefont{Bialek}},
  \bibinfo{author}{\bibfnamefont{A.}~\bibnamefont{Cavagna}},
  \bibinfo{author}{\bibfnamefont{I.}~\bibnamefont{Giardina}},
  \bibinfo{author}{\bibfnamefont{T.}~\bibnamefont{Mora}},
  \bibinfo{author}{\bibfnamefont{E.}~\bibnamefont{Silvestri}},
  \bibinfo{author}{\bibfnamefont{M.}~\bibnamefont{Viale}}, \bibnamefont{and}
  \bibinfo{author}{\bibfnamefont{A.~M.} \bibnamefont{Walczak}},
  \bibinfo{journal}{Proceedings of the National Academy of Sciences (USA)}
  \textbf{\bibinfo{volume}{109}}, \bibinfo{pages}{4786} (\bibinfo{year}{2012}).

\bibitem[{\citenamefont{Cavagna et~al.}(2015)\citenamefont{Cavagna,
  Del~Castello, Dey, Giardina, Melillo, Parisi, and Viale}}]{cavagna+al_2015}
\bibinfo{author}{\bibfnamefont{A.}~\bibnamefont{Cavagna}},
  \bibinfo{author}{\bibfnamefont{L.}~\bibnamefont{Del~Castello}},
  \bibinfo{author}{\bibfnamefont{S.}~\bibnamefont{Dey}},
  \bibinfo{author}{\bibfnamefont{I.}~\bibnamefont{Giardina}},
  \bibinfo{author}{\bibfnamefont{S.}~\bibnamefont{Melillo}},
  \bibinfo{author}{\bibfnamefont{L.}~\bibnamefont{Parisi}}, \bibnamefont{and}
  \bibinfo{author}{\bibfnamefont{M.}~\bibnamefont{Viale}},
  \bibinfo{journal}{Physical Review E} \textbf{\bibinfo{volume}{92}},
  \bibinfo{pages}{012705} (\bibinfo{year}{2015}).

\bibitem[{\citenamefont{Duncker and Sahani}(2021)}]{duncker2021dynamics}
\bibinfo{author}{\bibfnamefont{L.}~\bibnamefont{Duncker}} \bibnamefont{and}
  \bibinfo{author}{\bibfnamefont{M.}~\bibnamefont{Sahani}},
  \bibinfo{journal}{Current opinion in neurobiology}
  \textbf{\bibinfo{volume}{70}}, \bibinfo{pages}{163} (\bibinfo{year}{2021}).

\bibitem[{\citenamefont{Shenoy et~al.}(2013)\citenamefont{Shenoy, Sahani, and
  Churchland}}]{Shenoy2013}
\bibinfo{author}{\bibfnamefont{K.~V.} \bibnamefont{Shenoy}},
  \bibinfo{author}{\bibfnamefont{M.}~\bibnamefont{Sahani}}, \bibnamefont{and}
  \bibinfo{author}{\bibfnamefont{M.~M.} \bibnamefont{Churchland}},
  \bibinfo{journal}{Annual Review of Neuroscience}
  \textbf{\bibinfo{volume}{36}}, \bibinfo{pages}{337} (\bibinfo{year}{2013}).

\bibitem[{\citenamefont{Cunningham and Yu}(2014)}]{Cunningham2014}
\bibinfo{author}{\bibfnamefont{J.~P.} \bibnamefont{Cunningham}}
  \bibnamefont{and} \bibinfo{author}{\bibfnamefont{B.~M.} \bibnamefont{Yu}},
  \bibinfo{journal}{Nature Neuroscience} \textbf{\bibinfo{volume}{17}},
  \bibinfo{pages}{1500} (\bibinfo{year}{2014}).

\bibitem[{\citenamefont{Stringer et~al.}(2019)\citenamefont{Stringer,
  Pachitariu, Steinmetz, Carandini, and Harris}}]{Carsen2019}
\bibinfo{author}{\bibfnamefont{C.}~\bibnamefont{Stringer}},
  \bibinfo{author}{\bibfnamefont{M.}~\bibnamefont{Pachitariu}},
  \bibinfo{author}{\bibfnamefont{N.}~\bibnamefont{Steinmetz}},
  \bibinfo{author}{\bibfnamefont{M.}~\bibnamefont{Carandini}},
  \bibnamefont{and} \bibinfo{author}{\bibfnamefont{K.~D.}
  \bibnamefont{Harris}}, \bibinfo{journal}{Nature}
  \textbf{\bibinfo{volume}{571}}, \bibinfo{pages}{361} (\bibinfo{year}{2019}).

\bibitem[{\citenamefont{Nieh et~al.}(2021)\citenamefont{Nieh, Schottdorf,
  Freeman, Low, Lewallen, Koay, Pinto, Gauthier, Brody, and Tank}}]{Edward2021}
\bibinfo{author}{\bibfnamefont{E.~H.} \bibnamefont{Nieh}},
  \bibinfo{author}{\bibfnamefont{M.}~\bibnamefont{Schottdorf}},
  \bibinfo{author}{\bibfnamefont{N.~W.} \bibnamefont{Freeman}},
  \bibinfo{author}{\bibfnamefont{R.~J.} \bibnamefont{Low}},
  \bibinfo{author}{\bibfnamefont{S.}~\bibnamefont{Lewallen}},
  \bibinfo{author}{\bibfnamefont{S.~A.} \bibnamefont{Koay}},
  \bibinfo{author}{\bibfnamefont{L.}~\bibnamefont{Pinto}},
  \bibinfo{author}{\bibfnamefont{J.~L.} \bibnamefont{Gauthier}},
  \bibinfo{author}{\bibfnamefont{C.~D.} \bibnamefont{Brody}}, \bibnamefont{and}
  \bibinfo{author}{\bibfnamefont{D.~W.} \bibnamefont{Tank}},
  \bibinfo{journal}{Nature} \textbf{\bibinfo{volume}{595}}, \bibinfo{pages}{80}
  (\bibinfo{year}{2021}).

\bibitem[{\citenamefont{Recanatesi et~al.}(2021)\citenamefont{Recanatesi,
  Farrell, Lajoie, Den\`eve, Rigotti, and Shea-Brown}}]{Recanatesi2021}
\bibinfo{author}{\bibfnamefont{S.}~\bibnamefont{Recanatesi}},
  \bibinfo{author}{\bibfnamefont{M.}~\bibnamefont{Farrell}},
  \bibinfo{author}{\bibfnamefont{G.}~\bibnamefont{Lajoie}},
  \bibinfo{author}{\bibfnamefont{S.}~\bibnamefont{Den\`eve}},
  \bibinfo{author}{\bibfnamefont{M.}~\bibnamefont{Rigotti}}, \bibnamefont{and}
  \bibinfo{author}{\bibfnamefont{E.}~\bibnamefont{Shea-Brown}},
  \bibinfo{journal}{Nature Communications} \textbf{\bibinfo{volume}{12}},
  \bibinfo{pages}{1417} (\bibinfo{year}{2021}).

\bibitem[{\citenamefont{Cocco et~al.}(2011)\citenamefont{Cocco, Monasson, and
  Sessak}}]{Cocco2011}
\bibinfo{author}{\bibfnamefont{S.}~\bibnamefont{Cocco}},
  \bibinfo{author}{\bibfnamefont{R.}~\bibnamefont{Monasson}}, \bibnamefont{and}
  \bibinfo{author}{\bibfnamefont{V.}~\bibnamefont{Sessak}},
  \bibinfo{journal}{Physical Review E} \textbf{\bibinfo{volume}{83}}
  (\bibinfo{year}{2011}).

\bibitem[{\citenamefont{Krotov and Hopfield}(2016)}]{krotov+hopfield_2016}
\bibinfo{author}{\bibfnamefont{D.}~\bibnamefont{Krotov}} \bibnamefont{and}
  \bibinfo{author}{\bibfnamefont{J.~J.} \bibnamefont{Hopfield}}, in
  \emph{\bibinfo{booktitle}{Advances in Neural Information Processing
  Systems}}, edited by \bibinfo{editor}{\bibfnamefont{D.}~\bibnamefont{Lee}},
  \bibinfo{editor}{\bibfnamefont{M.}~\bibnamefont{Sugiyama}},
  \bibinfo{editor}{\bibfnamefont{U.}~\bibnamefont{Luxburg}},
  \bibinfo{editor}{\bibfnamefont{I.}~\bibnamefont{Guyon}}, \bibnamefont{and}
  \bibinfo{editor}{\bibfnamefont{R.}~\bibnamefont{Garnett}}
  (\bibinfo{publisher}{Curran Associates, Inc.}, \bibinfo{year}{2016}),
  vol.~\bibinfo{volume}{29}, pp. \bibinfo{pages}{1172--1180}.

\bibitem[{\citenamefont{Parisi}(1988)}]{parisi1988statistical}
\bibinfo{author}{\bibfnamefont{G.}~\bibnamefont{Parisi}},
  \emph{\bibinfo{title}{Statistical Field Theory}}
  (\bibinfo{publisher}{Frontiers in Physics, Addison-Wesley},
  \bibinfo{year}{1988}).

\bibitem[{\citenamefont{Sethna}(2021)}]{sethna2021statistical}
\bibinfo{author}{\bibfnamefont{J.}~\bibnamefont{Sethna}},
  \emph{\bibinfo{title}{Statistical Mechanics: Entropy, Order Parameters, and
  Complexity}} (\bibinfo{publisher}{Oxford University Press, Oxford},
  \bibinfo{year}{2021}).

\bibitem[{\citenamefont{Tka{\v{c}}ik et~al.}(2014)\citenamefont{Tka{\v{c}}ik,
  Marre, Amodei, Schneidman, Bialek, and Berry}}]{tkavcik2014searching}
\bibinfo{author}{\bibfnamefont{G.}~\bibnamefont{Tka{\v{c}}ik}},
  \bibinfo{author}{\bibfnamefont{O.}~\bibnamefont{Marre}},
  \bibinfo{author}{\bibfnamefont{D.}~\bibnamefont{Amodei}},
  \bibinfo{author}{\bibfnamefont{E.}~\bibnamefont{Schneidman}},
  \bibinfo{author}{\bibfnamefont{W.}~\bibnamefont{Bialek}}, \bibnamefont{and}
  \bibinfo{author}{\bibfnamefont{M.~J.} \bibnamefont{Berry}},
  \bibinfo{journal}{PLoS Computational Biology} \textbf{\bibinfo{volume}{10}},
  \bibinfo{pages}{e1003408} (\bibinfo{year}{2014}).

\bibitem[{\citenamefont{Meshulam et~al.}(2019)\citenamefont{Meshulam, Gauthier,
  Brody, Tank, and Bialek}}]{meshulam2019RG}
\bibinfo{author}{\bibfnamefont{L.}~\bibnamefont{Meshulam}},
  \bibinfo{author}{\bibfnamefont{J.~L.} \bibnamefont{Gauthier}},
  \bibinfo{author}{\bibfnamefont{C.~D.} \bibnamefont{Brody}},
  \bibinfo{author}{\bibfnamefont{D.~W.} \bibnamefont{Tank}}, \bibnamefont{and}
  \bibinfo{author}{\bibfnamefont{W.}~\bibnamefont{Bialek}},
  \bibinfo{journal}{Physical Review Letters} \textbf{\bibinfo{volume}{123}},
  \bibinfo{pages}{178103} (\bibinfo{year}{2019}).

\bibitem[{\citenamefont{Dowson and Wragg}(1973)}]{DowsonD.1973}
\bibinfo{author}{\bibfnamefont{D.}~\bibnamefont{Dowson}} \bibnamefont{and}
  \bibinfo{author}{\bibfnamefont{A.}~\bibnamefont{Wragg}},
  \bibinfo{journal}{IEEE Transactions on Information Theory}
  \textbf{\bibinfo{volume}{19}}, \bibinfo{pages}{689} (\bibinfo{year}{1973}).

\bibitem[{\citenamefont{Tagliani}(2003)}]{Tagliani2003}
\bibinfo{author}{\bibfnamefont{A.}~\bibnamefont{Tagliani}},
  \bibinfo{journal}{Applied Mathematics Letters} \textbf{\bibinfo{volume}{16}},
  \bibinfo{pages}{519} (\bibinfo{year}{2003}).

\bibitem[{\citenamefont{Novi~Inverardi and Tagliani}(2021)}]{Inverardi2021}
\bibinfo{author}{\bibfnamefont{P.~L.} \bibnamefont{Novi~Inverardi}}
  \bibnamefont{and} \bibinfo{author}{\bibfnamefont{A.}~\bibnamefont{Tagliani}},
  \bibinfo{journal}{Mathematics} \textbf{\bibinfo{volume}{9}}
  (\bibinfo{year}{2021}).

\bibitem[{\citenamefont{Evans}(2022)}]{evans2022partial}
\bibinfo{author}{\bibfnamefont{L.~C.} \bibnamefont{Evans}},
  \emph{\bibinfo{title}{Partial differential equations}},
  vol.~\bibinfo{volume}{19} (\bibinfo{publisher}{American Mathematical
  Society}, \bibinfo{year}{2022}).

\bibitem[{\citenamefont{Bender and Orszag}(2013)}]{bender2013advanced}
\bibinfo{author}{\bibfnamefont{C.~M.} \bibnamefont{Bender}} \bibnamefont{and}
  \bibinfo{author}{\bibfnamefont{S.~A.} \bibnamefont{Orszag}},
  \emph{\bibinfo{title}{Advanced Mathematical Methods for Scientists and
  Engineers. I: Asymptotic Methods and Perturbation Theory}}
  (\bibinfo{publisher}{Springer}, \bibinfo{address}{New York},
  \bibinfo{year}{2013}).

\bibitem[{\citenamefont{Meshulam et~al.}(2021)\citenamefont{Meshulam, Gauthier,
  Brody, Tank, and Bialek}}]{Meshulam2021}
\bibinfo{author}{\bibfnamefont{L.}~\bibnamefont{Meshulam}},
  \bibinfo{author}{\bibfnamefont{J.~L.} \bibnamefont{Gauthier}},
  \bibinfo{author}{\bibfnamefont{C.~D.} \bibnamefont{Brody}},
  \bibinfo{author}{\bibfnamefont{D.~W.} \bibnamefont{Tank}}, \bibnamefont{and}
  \bibinfo{author}{\bibfnamefont{W.}~\bibnamefont{Bialek}},
  \bibinfo{journal}{arXiv preprint arXiv:2112.14735}  (\bibinfo{year}{2021}).

\bibitem[{\citenamefont{Lynn et~al.}(2025{\natexlab{a}})\citenamefont{Lynn, Yu,
  Pang, Palmer, and Bialek}}]{Lynn2023}
\bibinfo{author}{\bibfnamefont{C.~W.} \bibnamefont{Lynn}},
  \bibinfo{author}{\bibfnamefont{Q.}~\bibnamefont{Yu}},
  \bibinfo{author}{\bibfnamefont{R.}~\bibnamefont{Pang}},
  \bibinfo{author}{\bibfnamefont{S.~E.} \bibnamefont{Palmer}},
  \bibnamefont{and} \bibinfo{author}{\bibfnamefont{W.}~\bibnamefont{Bialek}},
  \bibinfo{journal}{Physical Review E} \textbf{\bibinfo{volume}{111}},
  \bibinfo{pages}{054411} (\bibinfo{year}{2025}{\natexlab{a}}).

\bibitem[{\citenamefont{Zhu et~al.}(1997)\citenamefont{Zhu, Wu, and
  Mumford}}]{zhu+al_1997}
\bibinfo{author}{\bibfnamefont{S.~C.} \bibnamefont{Zhu}},
  \bibinfo{author}{\bibfnamefont{Y.~N.} \bibnamefont{Wu}}, \bibnamefont{and}
  \bibinfo{author}{\bibfnamefont{D.}~\bibnamefont{Mumford}},
  \bibinfo{journal}{Neural Computation} \textbf{\bibinfo{volume}{9}},
  \bibinfo{pages}{1627} (\bibinfo{year}{1997}).

\bibitem[{\citenamefont{Grend\'ar~Jr and Grend\'ar}(2001)}]{MEMysteries}
\bibinfo{author}{\bibfnamefont{M.}~\bibnamefont{Grend\'ar~Jr}}
  \bibnamefont{and}
  \bibinfo{author}{\bibfnamefont{M.}~\bibnamefont{Grend\'ar}},
  \bibinfo{journal}{Entropy} \textbf{\bibinfo{volume}{3}}, \bibinfo{pages}{58}
  (\bibinfo{year}{2001}).

\bibitem[{\citenamefont{Grend\'ar~Jr and
  Grend{\'a}r}(2001)}]{grendar2001minimax}
\bibinfo{author}{\bibfnamefont{M.}~\bibnamefont{Grend\'ar~Jr}}
  \bibnamefont{and}
  \bibinfo{author}{\bibfnamefont{M.}~\bibnamefont{Grend{\'a}r}}, in
  \emph{\bibinfo{booktitle}{AIP Conference Proceedings}}
  (\bibinfo{organization}{American Institute of Physics},
  \bibinfo{year}{2001}), vol. \bibinfo{volume}{568}, pp.
  \bibinfo{pages}{49--60}.

\bibitem[{\citenamefont{Lynn et~al.}(2025{\natexlab{b}})\citenamefont{Lynn, Yu,
  Pang, Bialek, and Palmer}}]{Lynn+al_2025b}
\bibinfo{author}{\bibfnamefont{C.~W.} \bibnamefont{Lynn}},
  \bibinfo{author}{\bibfnamefont{Q.}~\bibnamefont{Yu}},
  \bibinfo{author}{\bibfnamefont{R.}~\bibnamefont{Pang}},
  \bibinfo{author}{\bibfnamefont{W.}~\bibnamefont{Bialek}}, \bibnamefont{and}
  \bibinfo{author}{\bibfnamefont{S.~E.} \bibnamefont{Palmer}},
  \bibinfo{journal}{Physical Review Research} \textbf{\bibinfo{volume}{7}},
  \bibinfo{pages}{L022039} (\bibinfo{year}{2025}{\natexlab{b}}).

\bibitem[{\citenamefont{Newman and Barkema}(1999)}]{newman1999monte}
\bibinfo{author}{\bibfnamefont{M.~E.~J.} \bibnamefont{Newman}}
  \bibnamefont{and} \bibinfo{author}{\bibfnamefont{G.~T.}
  \bibnamefont{Barkema}}, \emph{\bibinfo{title}{Monte Carlo Methods in
  Statistical Physics}} (\bibinfo{publisher}{Oxford University Press},
  \bibinfo{address}{Oxford}, \bibinfo{year}{1999}).

\bibitem[{\citenamefont{Carcamo et~al.}(2025)\citenamefont{Carcamo, Weaver,
  Dixit, and Lynn}}]{carcamo+al_2025}
\bibinfo{author}{\bibfnamefont{D.~P.} \bibnamefont{Carcamo}},
  \bibinfo{author}{\bibfnamefont{N.~J.} \bibnamefont{Weaver}},
  \bibinfo{author}{\bibfnamefont{P.~D.} \bibnamefont{Dixit}}, \bibnamefont{and}
  \bibinfo{author}{\bibfnamefont{C.~W.} \bibnamefont{Lynn}},
  \bibinfo{journal}{arXiv preprint arXiv:2505.01607}  (\bibinfo{year}{2025}).

\bibitem[{\citenamefont{Lang}(1987)}]{lang1987linear}
\bibinfo{author}{\bibfnamefont{S.}~\bibnamefont{Lang}},
  \emph{\bibinfo{title}{Linear Algebra}} (\bibinfo{publisher}{Springer},
  \bibinfo{address}{New York}, \bibinfo{year}{1987}).

\bibitem[{\citenamefont{Cocco et~al.}(2018)\citenamefont{Cocco, Monasson,
  Posani, Rosay, and Tubiana}}]{cocco2018statistical}
\bibinfo{author}{\bibfnamefont{S.}~\bibnamefont{Cocco}},
  \bibinfo{author}{\bibfnamefont{R.}~\bibnamefont{Monasson}},
  \bibinfo{author}{\bibfnamefont{L.}~\bibnamefont{Posani}},
  \bibinfo{author}{\bibfnamefont{S.}~\bibnamefont{Rosay}}, \bibnamefont{and}
  \bibinfo{author}{\bibfnamefont{J.}~\bibnamefont{Tubiana}},
  \bibinfo{journal}{Physica A} \textbf{\bibinfo{volume}{504}},
  \bibinfo{pages}{45} (\bibinfo{year}{2018}).

\bibitem[{\citenamefont{Bialek}(2012)}]{bialek2012biophysics}
\bibinfo{author}{\bibfnamefont{W.}~\bibnamefont{Bialek}},
  \emph{\bibinfo{title}{Biophysics: searching for principles}}
  (\bibinfo{publisher}{Princeton University Press}, \bibinfo{year}{2012}).

\bibitem[{\citenamefont{Shannon}(1948)}]{shannon1948mathematical}
\bibinfo{author}{\bibfnamefont{C.~E.} \bibnamefont{Shannon}},
  \bibinfo{journal}{The Bell System Technical Journal}
  \textbf{\bibinfo{volume}{27}}, \bibinfo{pages}{379} (\bibinfo{year}{1948}).

\bibitem[{\citenamefont{Cover and Thomas}(1991)}]{cover+thomas_91}
\bibinfo{author}{\bibfnamefont{T.~M.} \bibnamefont{Cover}} \bibnamefont{and}
  \bibinfo{author}{\bibfnamefont{J.~A.} \bibnamefont{Thomas}},
  \emph{\bibinfo{title}{Elements of Information Theory}}
  (\bibinfo{publisher}{Wiley and Sons, New York}, \bibinfo{year}{1991}).

\end{thebibliography}
\end{document}